\documentclass[iop]{emulateapj} 
\usepackage{amsmath}
\usepackage{amssymb}
\usepackage{mathbbol}
\usepackage{dsfont}

\DeclareGraphicsExtensions{.eps,.eps.gz,.epsi}

\newcommand{\teff}{\mbox{$T_{\rm eff}$}} \newcommand{\logg}{{\rm{log}~$g$}}
\newcommand{\feh}{{\rm [Fe/H]}} 
\newcommand{\ebv}{$E(B-V)$}

\shorttitle{Stellar color regression}
\shortauthors{Yuan et al.}

\begin{document}

\title
{Stellar color regression: a spectroscopy based method for color calibration to a few mmag accuracy and the recalibration of Stripe 82}

\author
{Haibo Yuan\altaffilmark{1, 2}, 
Xiaowei Liu\altaffilmark{3, 1}, 
Maosheng Xiang\altaffilmark{3},
Yang Huang\altaffilmark{3}, 
Huihua Zhang\altaffilmark{3}, 
Bingqiu Chen\altaffilmark{3}
}
\altaffiltext{1}{Kavli Institute for Astronomy and Astrophysics, Peking University, Beijing 100871, P. R. China; email: yuanhb4861@pku.edu.cn}
\altaffiltext{2}{LAMOST Fellow}
\altaffiltext{3}{Department of Astronomy, Peking University, Beijing 100871, P. R. China; email: x.liu@pku.edu.cn}

\journalinfo{submitted to The Astrophysical Journal}
\submitted{Received ; accepted }

\begin{abstract}
In this paper, we propose a spectroscopy based Stellar Color Regression (SCR) method to perform accurate color calibration for modern imaging surveys,
taking advantage of millions of stellar spectra now available.
The method is straightforward, insensitive to systematic errors in the spectroscopically determined stellar atmospheric parameters, 
applicable to regions that are effectively covered by spectroscopic surveys, 
and capable of delivering an accuracy of a few millimagnitudes for color calibration.
As an illustration, we have applied the method to the SDSS Stripe\,82 data (Ivezi{\'c} et al; I07 hereafter). 
With a total number of 23,759 spectroscopically targeted stars, we have mapped out the small but strongly correlated 
color zero point errors present in the photometric catalog of Stripe\,82, and improve 
the color calibration by a factor of 2 -- 3.
Our study also reveals some small but significant magnitude dependence errors in $z$-band for some CCDs. 
Such errors are likely to be present in all the SDSS photometric data. 
Our results are compared with those from a completely independent test based on the intrinsic colors of red galaxies presented by I07.
The comparison as well as other tests shows that the SCR method has achieved a color calibration internally consistent 
at a level of about 5 mmag in $u-g$,  3 mmag in $g-r$, and 2 mmag in $r-i$ and $i-z$, respectively.
Given the power of the SCR method, we discuss briefly the potential benefits by applying the method to existing, 
on-going, and up-coming imaging surveys.
\end{abstract}


\section{Introduction} 
Uniform and accurate photometric calibration plays a central role in the
current and next-generation wide-field imaging surveys, such as the Sloan
Digital Sky Survey (SDSS; York et al. 2000), the Dark Energy Survey (DES; The
DES Collaboration 2010), Pan-STARRS (Kaiser et al. 2002) and the Large Synoptic Survey
Telescope (LSST; Ivezi{\'c} et al. 2008b).  Accurate photometric calibrations over
wide areas are pivotal for robust object classifications, determinations of photometric 
redshifts of a  large number of galaxies to probe the large-scale structure of the universe 
(e.g. Padmanabhan et al. 2007; Blake et al. 2007), and for 
topographical studies of the Galactic structure via a large number of stars whose basic properties can be
obtained from the photometric magnitudes and colors (e.g. Juri{\'c} et al. 2008; Ivezi{\'c} et al. 2008a).

Astronomical optical photometric measurements are traditionally calibrated based on
sets of standard stars, such as the  Landolt's (Landolt 1992) and
Stetson's (Stetson 2000, 2005). 
However, achieving a one per cent precision for ground-based wide field imaging surveys is 
difficult given the spatial and temporal variations of the Earth 
atmospheric transmission and the instrumental effects such as flat-fielding  (e.g. Stubbs \& Tonry 2006). 

To pursuit a one per cent accuracy for large field imaging surveys, a number of different approaches have been proposed and implemented. 
One main approach is to decouple the problem of
``relative" calibrations (i.e., producing an internally consistent system) from
that of ``absolute" calibrations (i.e., connecting the internal system to a physical flux scale) 
using observations from the overlapping regions 
(e.g., Ivezi{\'c} et al. 2007, hereafter I07; Padmanabhan et al. 2008; Schlafly et al. 2012). 
The absolute calibration for the whole survey is then reduced to the determinations of a few numbers.
Applying the (ubercal) method to the whole SDSS imaging data, 
Padmanabhan et al. have achieved about 1 per cent relative calibration for $g, r, i$, and $z$ bands and
about 2 per cent for $u$ band. These errors are dominated by the unmodeled variations of atmospheric transmission.
Applying a similar method to the repeatedly scanned (at least four times per band, with a median of 10 times per band) 
SDSS Stripe 82 data, I07 have constructed a standard star catalog in the SDSS $ugriz$ system 
containing about 1 million objects with an internal photometric accuracy of 1 per cent 
[i.e. the spatial variations of the photometric zero points across the stripe survey area is smaller than 0.01\,mag (rms)].
The difference between the two approaches is that I07 use a different algorithm for flat-field corrections,
requiring that the stellar loci (Ivezi{\'c} et al. 2004; Covey et al. 2007) remain fixed in multi-dimensional color space,
whereas Padmanabhan et al. (2008) achieve so by minimizing the relative photometric errors of repeated observations. 

An alternative approach is to use the stellar locus regression method (SLR; High et al. 2009), which aims to achieve  
real-time, accurate color calibration using direct measurements from multi-band, flat-fielded images for a given field.
Note that colors hold the key information about a source's type, temperature, metallicity, and redshift.
Thus achieving accurate color calibration is often more important than deriving highly accurate (absolute) 
magnitudes in many astronomical applications, in particular considering the fact that 
we seldom know the distances to individual celestial objects to a per cent level.
The SLR method makes a wholesale
correction for a variety of potential effects, including variations in instrumental sensitivity, in the Earth atmospheric
extinction and reddening, and in the Galactic interstellar extinction and reddening
by adjusting the instrumental broadband optical colors of stars to bring them
into accord with a universal set of stellar color loci. 
The method is able to produce calibrated colors accurate to a few per cent.
The approach needs neither estimates of zero point for individual passbands nor repeated measurements of the standard stars, 
and can be performed in real-time as the survey progresses. 
However, the SLR method assumes the stellar loci are universal, an assumption that is not always true,
given the variations of stellar populations and the interstellar extinction, especially in the Galactic disk.
It also requires a blue filter in addition to at least two other filters.

In addition to imaging photometry for over 10,000 deg$^2$, the 
SDSS has delivered low-resolution spectra for about 0.7\,million stars in its Data Release 9 (DR9; Ahn et al. 2012).
By June 2014, the LAMOST Galactic surveys (Deng et al. 2012; Liu et al. 2014)
have collected over 3\,M stellar spectra and the stellar parameters (Bai et al. 2014, in preparation; Yuan et al., 2014a)
for over half of them. The surveys will eventually obtain over 6 million spectra upon the completion in three years' time. 
With millions of stellar spectra over large sky area now available, ``identical" stars 
located in different environments but otherwise having nearly identical atmospheric characteristics can be easily paired and compared
to measure the interstellar reddening of stars suffering from substantial extinction (Yuan, Liu \& Xiang 2013), 
as well as to detect and measure the strengths of the diffuse interstellar bands along hundreds of thousands sightlines (Yuan \& Liu 2012), 
assuming that stars of the same set of atmospheric parameters (\teff, \logg~and \feh) possess the same intrinsic colors and spectra. 
To take advantage of this era of millions of stellar spectra and the fact that stars observed spectroscopically serve as excellent color standards,
here we propose a spectroscopy based Stellar Color Regression (SCR) method for accurate relative color (re-)calibration of modern
large sky area wide field imaging surveys.

To apply the method, one requires that: 1) A few well-calibrated fields are available. Spectroscopic reference stars selected
from those defining fields are used to determine the stellar intrinsic colors as a function of the atmospheric parameters;
2) The reddening law does not vary within the sky area of a (target) field to be calibrated;
and 3) A sub-sample of stars in the field to be calibrated have been spectroscopically targeted and the
reddening values of those targets are known.
With those information, the zero points of colors can be derived by linear regression between the reddening and
the color offsets of stars of the target field relative to the intrinsic colors.
By this way, the whole survey fields can be brought on a uniform color scale.
Note that in cases where the reference fields have a calibration error, 
the method brings the photometry of all fields onto a uniform, internally consistent color scale that 
may have an overall offset from the true color scale. This is analogous to the Ubercalibration algorithm
(Padmanabhan et al. 2008) that brings the photometry of the whole survey footprint 
in each bandpass to a uniform scale that may have an overall offset from the AB system.
The method yields simultaneously the reddening coefficients as a by-product.
Compared to the SLR method, the SCR method has the advantages of being straightforward, 
model-free\footnote{The method is insensitive to possible systematic errors in the spectroscopically determined stellar atmospheric parameters.
It does assume that the reddening law does not vary within the area of a target field to be calibrated. 
However, as long as the target field is not very large or can be further divided into smaller sub-fields, 
depending on the source density, the assumption should be a reasonable one.}  
and applicable to large sky area encompassing a variety of environments (Galactic halo and disk, regions of low or high extinction) 
that are effectively covered by large-scale spectroscopic surveys such as the SDSS and LAMOST.
Given a spectroscopic sampling density of 
a few hundred stars per deg$^2$, and a precision of respectively 
30 -- 100\,K, 0.1 -- 0.25\,dex, and 0.05 -- 0.1\,dex for the spectroscopically yielded atmospheric parameters 
\teff, \logg~and \feh~(Lee et al. 2008a, 2008b; Wu et al. 2011; Xiang et al. 2014),
the SCR method is capable of delivering a few mmag accuracy for relative color calibration. 

To demonstrate the power and performance of the SCR method, 
we have applied the method to the Stripe 82 data of I07. 
As demonstrated by I07 with a variety of independent tests, the spatial variations of 
the photometric zero points must be no larger than 0.01 mag in $griz$ bands and 0.02 mag in $u$ band (rms). 
Applying the SCR method to a sample of 23,759 spectroscopically targeted stars
of Stripe\,82  in a sky area of  
about 80 per deg$^2$, we have confirmed the conclusions of I07. 
However, the study also reveals small yet substantial and tightly correlated 
spatial variations in the color zero points across the stripe, and are ascribed to 
the degeneracies intrinsic to the I07 calibration.
Possible color and magnitude dependence of the variations is also examined.
Small but significant magnitude dependence is found in $z$-band for some CCDs.
Our results are compared with those from a completely independent test based on the intrinsic colors of red galaxies.
The comparison confirms that the SCR method has achieved a
calibration accuracy of or better than 5 mmag in color $u-g$,  3 mmag in color $g-r$, and 2 mmag in colors $r-i$ and $i-z$, respectively.
The uncertainties are only a factor of two larger than what expected from the random errors alone. 
The method can thus serve as an extremely powerful and robust tools to 
calibrate existing, on-going, and up-coming large scale imaging surveys,  including the SDSS, DES, Pan-STARRS, and LSST.

The paper is organized as follows.  In Section 2, we introduce the method.
In Section\,3, the method is applied to the SDSS Stripe 82 region to demonstrate its performance and recalibrate the data. 
The discussions and conclusions are given in Section\,4.

\section{The SCR method} 
\begin{figure}
\includegraphics[width=90mm]{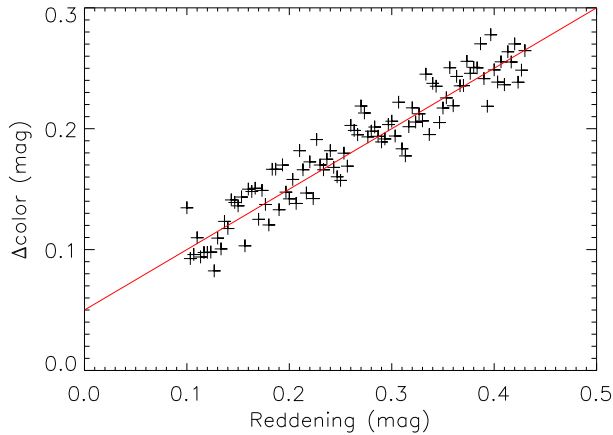}
\caption{The plot illustrates how the SCR method works.
For a given field to be calibrated against the reference field, the dots plot 
the differences between the measured and intrinsic colors of
stars spectroscopically targeted in the field against estimates of reddening. 
The line is a linear regression of the data points.
The slope and intercept give respectively 
the reddening coefficient and color zero point of the field.  
}
\label{}
\end{figure}

The current method is built on the spectroscopic ``star pair" technique (Yuan \& Liu 2012; Yuan, Liu \& Xiang 2013), 
taking advantage of millions of stellar spectra and parameters now available over a large sky area. 
The star pair technique has been used to estimate multi-band interstellar reddening and reddening coefficients 
of a large number of stars targeted by the SDSS (Yuan, Liu \& Xiang 2013) and by the LAMOST (Yuan et al. 2014a), and 
to detect and measure the diffuse interstellar bands in thousands of the SDSS spectra (Yuan \& Liu 2012).
Unlike in the above work where star pairs are composed of target and control stars from high and low extinction 
regions, respectively, in the current work they are composed of target and control stars from the fields to be calibrated and 
from those of well calibrated, reference fields that are used to define 
the stellar intrinsic colors as a function of stellar atmospheric parameters. 
For each star in the target sample, its control stars are selected from the control sample as
those having values of \teff, \logg~and \feh~that differ from those of
the target by less than 200 K, 0.25 dex, and 0.5 dex, respectively. 
The intrinsic colors of the target are derived by linearly interpolating  
the intrinsic colors of its control stars in \teff~and \feh. 
Linear interpolation should be sufficient given 
the small ranges of values of parameters being considered. The dependence of colors on \logg~has been ignored here. 
Stars with less than three control stars are not used.

This method requires that 
a sub-sample of target stars in a given field to be calibrated have been spectroscopically observed and their 
reddening values in a certain color are available,
say from the 2D extinction map of Schlegel et al. (1998; SFD hereafter) for targets at 
high Galactic latitudes or derived using the star pair technique of Yuan, Liu \& Xiang (2013)  
from existing and well calibrated photometric surveys for targets at low Galactic latitudes. 
The reddening law should also not vary across the field to be calibrated 
(thus the field should not be too large and encompass different environments).
In general, the reddening law is expected to show no or only marginal spatial variations at 
high Galactic latitudes. In the Galactic disk, 
it is however known to vary significantly and data to constrain such variations are now available (e.g. Yuan et al. 2014a). 
Finally to obtain a highly accurate definition of the stellar intrinsic colors as a function of atmospheric parameters, 
one or a few photometrically and spectroscopically well-calibrated fields are needed.
Those fields will be used to determine the intrinsic colors from a given set of stellar parameters.
Then for a given target field to be calibrated, 
the method obtains its color zero point by performing a linear regression between the reddening and
color offsets relative to the intrinsic colors using the sub-sample of spectroscopically targeted stars of the field.
Calibrating each target field by this way, this method will bring the whole survey on a uniform color scale.
This process will also yield the reddening coefficients of each target field as a by-product.
Fig.\,1 schematically illustrates the method.
Note that if the reddening coefficient is previously known, then the slope of regression can be 
fixed when fitting the data. 

The accuracy achievable with the method is controlled by:
1) the random errors of color measurements; 2) the random errors of atmospheric parameters of the 
reference stars as well as of the spectroscopically targeted sub-sample of stars in the fields to be calibrated;
3) the random errors and breadth of range spread by the reddening values of the spectroscopic sub-sample of target stars;  
and 4) the size of the spectroscopic sub-sample of target stars in the individual target fields.
For modern digital imaging surveys, the random errors of measured colors are typically at 1 -- 2 per cent level
at good signal-to-noise ratios (SNRs). The random errors of stellar atmospheric parameters 
derivable from low- and medium resolution ($R\sim 2,000$) spectra of SNRs better than 20 
with those widely used stellar parameter pipelines [such as the SSPP -- the Sloan Extension for Galactic 
Understanding and Exploration (SEGUE; Yanny et al. 2009) Stellar Parameter Pipeline (Lee et al. 2008a,b), 
or the LSP3 -- the LAMOST Stellar Parameter Pipeline at Peking University (Xiang et al. 2014)]
are at a level of 0.005 -- 0.01 dex (i.e., about 75 -- 150\,K at 6,000 K) in \teff~
and 0.05 -- 0.1\,dex in \feh. Systematic errors in the stellar atmospheric parameters 
do not affect the performance of the SCR method, as they are cancelled out. 
Taking the SDSS colors for example, a 0.01 dex change in \teff~for FGK stars will lead to 
variations of 0.08, 0.04, 0.013, and 0.010 mag in colors $u-g$, $g-r$, $r-i$ and $i-z$, respectively. 
Similarly, a 0.1 dex change in \feh~for FGK stars can cause about variations of 
0.05, 0.023, and 0.01 mag in color $u-g$ at \feh~=~0, $-1$, and  $-2$\,dex, respectively,
and essentially nil ($<$ 0.015\,mag) in the other three colors. 

At high Galactic latitudes, reddening values of the target stars can be obtained from the SFD extinction map,
or the Planck extinction map (Planck Collaboration et al. 2013). 
The random errors of the SFD extinction are very small, although their values are known to 
suffer from some systematic errors due to a scaling problem 
(e.g. Schlafly et al. 2010; Schlafly \& Finkbeiner 2011; Yuan, Liu \& Xiang 2013). 
In the SCR method, those systematics however can be taken care of 
by the reddening coefficients, and thus should not affect its results.
At low Galactic latitudes where the SFD extinction map is not usable, 
reddening values of the target stars can be derived using the star pair technique (Yuan, Liu \& Xiang 2013)
or, alternatively, the spectral energy distribution (SED) fitting method (e.g. Berry et al. 2012). 
For example, using the SED fitting method, Chen et al. (2014) have determined the 
reddening values for over 13 million stars within the footprint of the 
Xuyi Schmidt Telescope Photometric Survey of the Galactic Anti-centre (XSTPS-GAC; Liu et al. 2014)
by combining the XSTPS-GAC optical photometry and the near infrared photometry of 2MASS (Skrutskie et al. 2006) 
and the Wide-field Infrared Survey Explorer (WISE; Wright et al. 2010). For objects targeted by the 
LAMOST Spectroscopic Survey of the Galactic Anti-centre (LSS-GAC; Liu et al. 2014), over a million already, 
reddening values have also been deduced using the star pair method and 
released in the value-added catalogs (Yuan et al. 2014a) supplementary to the LAMOST official data release.  
The random errors of reddening values thus obtained after converted into \ebv~are typically between 0.02 to 0.1 mag, 
depending on which datasets are available. 
There are however some degeneracies between the color calibration and reddening coefficient determinations.
If the stars in a given target field span a wide range of reddening values (relative to the uncertainties of reddening), 
which is likely the case for high extinction regions of low Galactic latitudes, the reddening coefficients and 
the color zero points can be robustly determined. 
In cases where the reddening values of the target stars are large yet span a narrow range,
incorporating a prior distribution of reddening coefficients with a Bayesian approach may be needed 
to improve the accuracy of color calibration.
For high Galactic latitude fields where the reddening values of target stars are small and span a narrow range,
fixing the reddening coefficients to some pre-determined values is probably favored.
The number of spectroscopically targeted stars is about one hundred to a few hundred per deg$^2$, as afforded by surveys 
such as the SDSS and LAMOST. 
Here we have neglected the random errors of reddening of the reference stars 
considering that the reference stars are usually selected from well-calibrated, low extinction regions 
and thus suffer from negligible uncertainties in reddening corrections. 
Given the above constrains, a few per cent accuracy in color measurements is capable of 
delivering a few mmag accuracy in color calibration with the SCR method, as we shall show in the next Section.

\section{Recalibration of Stripe 82}

The SCR method can be used to check the calibration accuracies of existing photometric data and recalibrate them.
Any systematic photometric errors present in the calibrated data, except for grey ones that are the same in different bands,
will reveal themselves as offsets in dereddened colors with respect to the intrinsic values. 
In this section, we will apply the method to the SDSS Stripe 82  
to check the accuracy of the current calibration of I07 and further recalibrate the data.

\subsection{Data}
\begin{figure}
\includegraphics[width=90mm]{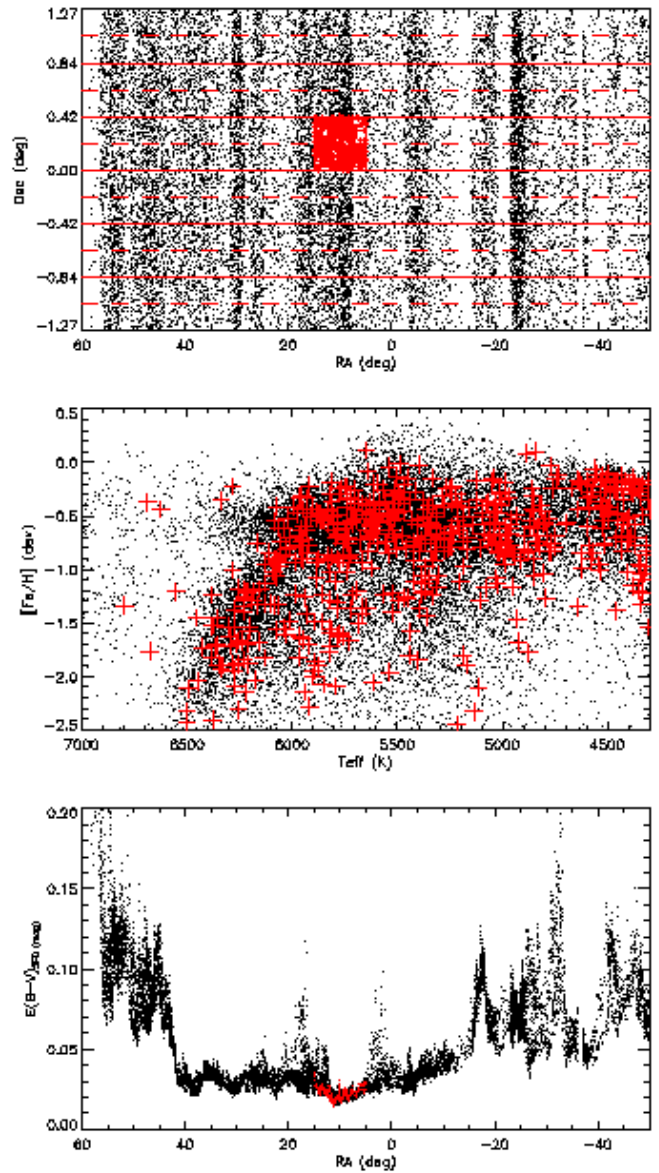}
\caption{Distributions of the spectroscopic sample of stars in the RA -- Dec (top), \teff~-- \feh~(middle) 
and RA -- \ebv~(bottom) planes. Control stars from the reference field 
are indicated by red pluses. The centres of individual scans (scan lines) 
are marked by red horizontal dashed lines.
}
\label{}
\end{figure}

The repeatedly scanned equatorial Stripe 82 ($|{\rm Dec}| < 1.266\degr$, 20$^h$34$^m$ $<$ RA $<$ 4$^h$00$^m$)
in the SDSS has delivered the most accurate photometric catalog of 
about one million stars in $u,g,r,i,z$ bands (I07).
This is the largest optical photometric data set publicly available with an internal calibration consistency of one per cent,
thus providing ``a practical definition of the SDSS photometric system" (I07).
The random photometric errors of the I07 catalog are below 0.01\,mag for stars brighter 
than 19.5, 20.5, 20.5, 20.0, and 18.5\,mag in $u$, $g$, $r$, $i$, and $z$ band, respectively, 
about twice as good as for the individual SDSS runs.
Within the Stripe, over 40,000 stellar spectra have been collected and released 
in the SDSS DR9 (Ahn et al. 2012),
along with the basic stellar parameters derived from the spectra with the SSPP pipeline. 

Note that one SDSS stripe consists of two drift scan strips. Each strip is scanned by six 
columns of five rows of CCDs with gaps between the columns. 
Please see Fig.\,1 of Padmanabhan et al. (2008) for an illustration of the geometry of SDSS imaging.
Each CCD covers a sky area of 13 arcmin 
perpendicular to the scan direction. 
The two strips, offset by 12.6\,arcmin (thus an overlap of about 1\,arcmin for each CCD), 
form a stripe of contiguous sky coverage of width 150\,arcmin. 
The five rows of CCDs in the scan direction, are filtered with filters of  $r$, $i$, $u$, $z$, and $g$, respectively. 
As the telescope scans, photometry in the above 5 bands are collected in sequence (cf. York et al. 2000).

We first select stars of Stripe 82 from the SDSS DR9 
that are spectroscopically targeted and have 
a spectral SNR better than 20 and an effective temperature \teff~between 4,300 -- 7,000\,K.
Stars hotter than 7,000\,K or cooler than 4,300\,K are excluded since their atmospheric parameters are less 
reliable\footnote{We have used the ``adopted" atmospheric parameters recommended by the SSPP.
The SSPP determines effective temperatures using a variety of methods, some based on photometric colors.
A comparison of the SSPP ``adopted'' values with those derived based on spectroscopic measurements only
yields a scatter of about 30\,K in \teff~and 0.05 dex in \feh.
Replacing the ``adopted'' values with those based on spectroscopy only has 
little effects on the results presented here.}.
Then we cross-match this sample with the I07 catalog, 
with a match radius of 1 arcsec. In total, 23,759 stars are selected.
Then a sub-sample of 574 stars in a small area of 5 $<$ RA $<$ 15\,deg and 0 $<$ Dec $<$ 0.422\,deg 
and of line-of-sight extinction values \ebv~from the SFD map less than 0.04\,mag is further selected as the control stars. 
The control stars are all recorded with the same CCD. 
This control sample is used as a reference to define the intrinsic colors as a function of the atmospheric parameters, 
which in turn are used to map the spatial variations of offset of color zero point for all other fields.
The mean magnitudes given by the I07 catalog are used.
The distributions of the selected stars spectroscopically targeted, including the sub-sample of control stars, 
in the RA -- Dec, \teff~-- \feh~and ${\rm RA}$ -- \ebv~planes are displayed in Fig.\,2.
Note that the sample stars are not distributed homogeneously in RA.

\subsection{Reddening corrections and intrinsic colors} 
\begin{figure*}
\includegraphics[width=180mm]{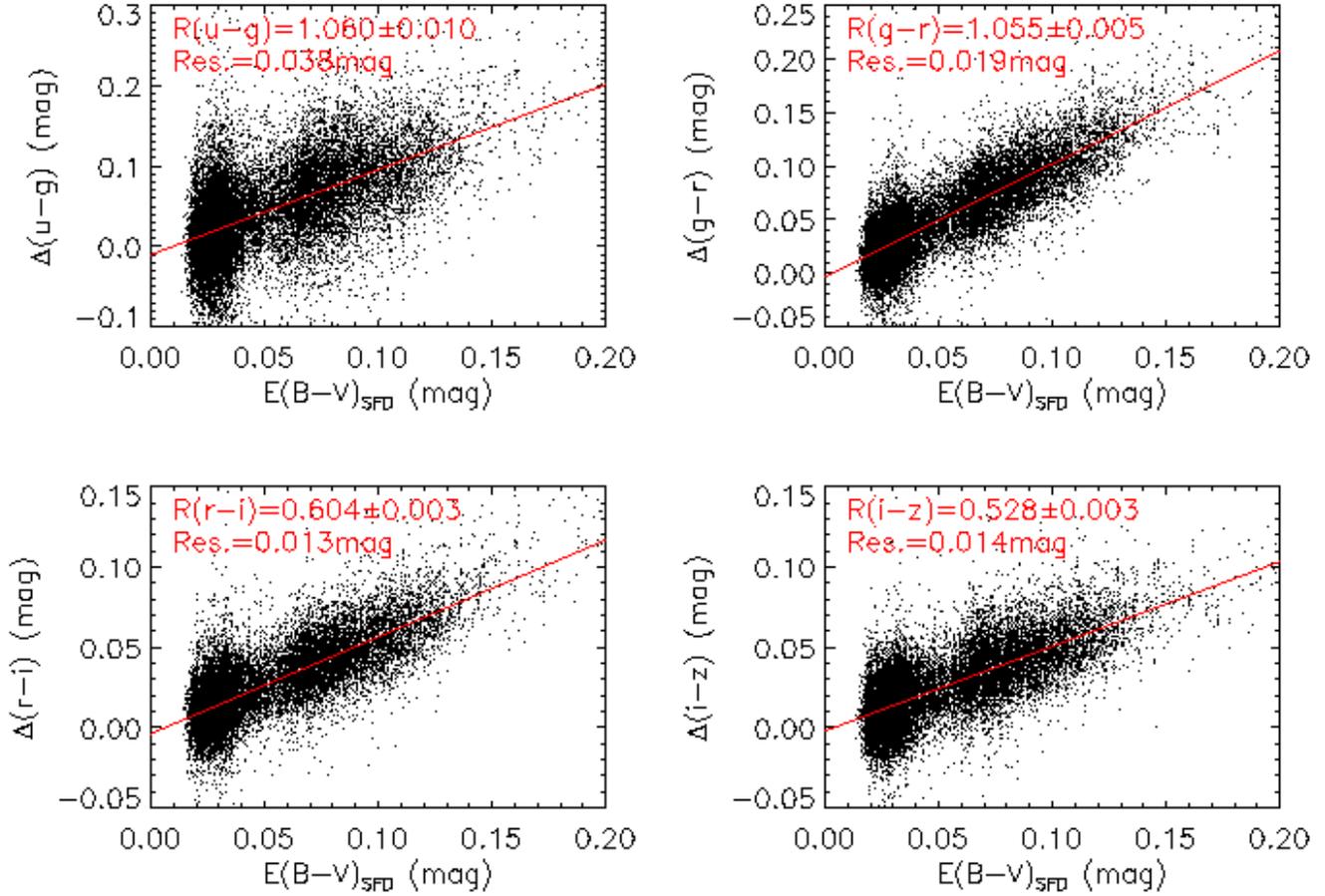}
\caption{
Values of reddening in $u-g$, $g-r$, $r-i$, and $i-z$ colors plotted against \ebv~from the 
SFD extinction map for the spectroscopic sample.
The red lines are linear fits to the data points. The resultant reddening coefficients and errors,
and the fit residuals are labelled. 
}
\label{}
\end{figure*}

Given that the Stripe 82 is located at high Galactic latitudes, the SFD extinction map is used 
to assign reddening values to all spectroscopically targeted stars, including the sub-sample of control stars. 
The reddening values of target stars within a limited area of Stripe 82 usually span only a narrow range (Fig.\,2).
In order to avoid possible degeneracies between the color calibration and reddening coefficient determinations,
when applying the SCR method to Stripe\,82, 
we have neglected any possible spatial variations of reddening coefficients across the Stripe induced by 
changes in reddening law or systematics in the SFD map. 
Given the high Galactic latitudes 
and small values of reddening of the Stripe, 
the impacts of this simplified approach on the results are likely to be small. 
For example, a ten per cent variation in the reddening coefficients, which is highly unlikely, 
will result in uncertainties in the reddening corrections and color calibrations of 
less than a few mmag only, given the very small values of reddening.

To determine the reddening coefficients in colors $u-g$, $g-r$, $r-i$, and $i-z$ for the whole spectroscopic sample of Stripe 82,
we have followed the procedures of Yuan, Liu \& Xiang (2013). 
The sub-sample of control stars are first dereddened using an initial set of reddening coefficients and \ebv~values from SFD. 
The values of color excess of the whole spectroscopic sample of stars 
are then estimated by comparing their observed colors with the intrinsic ones, 
with the latter derived using the star pair technique described in Section\,2.   
Then a new set of reddening coefficients is derived by linear regression between the estimates of color excess and values of \ebv.  
No passing through the origin is forced in the regression. 
Iterations are carried out till the derived set of reddening coefficients is 
consistent with the one used to deredden the control sample.
The results are shown in Fig.\,3. 
It is found that $R_{u-g}$ = $1.056\pm0.009$, $R_{g-r}$ = $1.050\pm0.004$, $R_{r-i}$ = $0.599\pm0.003$, and $R_{i-z}$ = $0.523\pm0.003$. 
The results suggest that the SFD map overestimates the true reddening values by about 8 per cent for Stripe 82. 
The small errors of the derived reddening coefficients confirm that the spatial 
variations of the reddening coefficients across Stripe\,82 are insignificant if any, as assumed. 
The small scatters of the fit residuals, 0.038, 0.019, 0.013, and 0.014\,mag for colors
$u-g$, $g-r$, $r-i$, and $i-z$, respectively, suggest that: 
1) The SFD extinction map traces the interstellar dust reddening well at high Galactic latitudes;
2) The random errors of the SSPP adopted atmospheric parameters are small; 
and 3) The intrinsic colors (and values of color excess) of a given star in the whole spectroscopic sample 
can be determined to an accuracy of about 0.04\,mag in $u-g$ and about 0.01 -- 0.02\,mag in other colors.

Using dwarf stars of the control sample, we have performed two dimensional polynomial fitting          
to their dereddened $u-g$, $g-r$, $r-i$, and $i-z$ colors as a function of \teff~and \feh.
For color $u-g$, a 4th-order polynomial with 15 free parameters is adopted.
For colors $g-r$, $r-i$, and $i-z$, a 3rd-order polynomial with 10 free parameters is used.
The fit residuals are 0.032, 0.017, 0.013, and 0.012\,mag for colors
$u-g$, $g-r$, $r-i$, and $i-z$, respectively. The numbers are slightly smaller 
than the fit residuals in Fig.\,3, suggesting that the reduced $\chi^2$ of the fits in Fig.\,3 are 
slightly larger than, but close to, unity.
In other words, the scatters in the fits of Fig.\,3 are mainly contributed by photometric errors,          
random errors in \teff~ and \feh, and calibration errors.
The small uncertainties in the recovered reddening coefficients, maybe under-estimated slightly, are due to
large numbers of stars used and accurate estimates of \ebv~of individual stars.
We also divide the whole spectroscopic sample into six bins in RA to derive reddening coefficients seperately. 
For three of them, the coverage in \ebv~is too small to obtain a valid fit. For the other three bins,                 
the yielded reddening coefficients are consistent with each other within 2.5-sigma uncertainties.

With the reddening coefficients determined above,
the whole spectroscopic sample of stars, including the
control stars from the reference field, are dereddened using the SFD map. 
From the dereddened colors of the control stars, 
the intrinsic colors of the whole spectroscopic sample of stars are also estimated 
using the star pair technique described in Section\,2.
Given that we have fixed the reddening coefficients for the whole field of Stripe 82,
the (error of) color zero point for a given field to be (re-)calibrated can be determined directly by 
comparing the intrinsic and dereddened colors of stars spectroscopically targeted in the field.
However, unlike most imaging surveys where the color zero points refer to the individual fields,
the SDSS imaging data are collected in drift-scan mode, thus in principle 
there are no traditionally defined fields for the survey.
The zero points for one SDSS strip are continuous functions of the spatial position (observational time) 
of the data points along the great circle of the scan.
In the case of Stripe 82, the systematic errors of color calibration are functions of RA and Dec.
Once the intrinsic and dereddened colors for the whole sample of spectroscopically targeted stars are obtained, 
the offsets in each of the $u-g$, $g-r$, $r-i$, and $i-z$ colors for each star are estimated. 
The variations of color offsets as a function of RA and Dec as well as their possible dependence on colors and magnitudes
are investigated in the following subsections to check the calibration of I07 and recalibrate the data.

\subsection{Spatial variations of color offsets in RA $\delta_c^{ext}({\rm RA})$}
Following I07, the true colors of an object, $c_{\rm true}$, can be expressed as 
\begin{equation} 
c_{true} = c_{cat} + \delta_c^{ext}({\rm RA}) + \delta_c^{ff}({\rm Dec}),
\end{equation}
where $c_{\rm cat}$ is the cataloged magnitude,
$\delta_c^{ext}({\rm RA})$ is dominated by unrecognized fast variations of the atmospheric extinction, 
$\delta_c^{ff}({\rm Dec})$ is dominated by errors in the flat-field vectors. 
In the current work, $\delta_c^{ext}({\rm RA})$ and $\delta_c^{ff}({\rm Dec})$ arise from uncorrected for 
or over-corrected errors in I07. Note that in this work, we do not 
consider any systematic color errors that might be present in the whole catalog (i.e., assume that the overall systematic color
errors are zero) but aim to investigate the possible internal (spatial) variations of the color calibration.
Thus, the average of $\delta_c$ over the whole cataloged area is zero by definition.
Here we have also neglected systematic effects such as the device non-linearity and 
the bandpass variations between the different camera columns, which depend on the properties of individual sources such as 
brightness and color. We will however explore those effects in subsection\,3.5. 
In this subsection, we will first investigate the spatial variations of color offsets in RA, i.e., $\delta_c^{ext}({\rm RA})$.

Stripe 82 is an elongated rectangle area of an aspect ratio of 1:50, with the long side parallel to the celestial equator. 
We divide the Stripe into 100 bins of RA, with each bin containing an equal number (about 200) of (spectroscopically targeted) stars.
For each bin, the median values of RA and color offsets in the $u-g$, $g-r$, $r-i$, and $i-z$ colors are estimated.
The results are listed in Table\,2 of the Appendix and plotted in Fig.\,4.
Given the number of stars in each bin (about 200) and the typical errors of color offsets for a single star 
are about 0.039, 0.019, 0.013, and 0.013 mag for the $u-g$, $g-r$, $r-i$, and $i-z$ colors,  
the median color offsets of each bin have errors of about 2.8, 1.3, 0.9, and 0.9 mmag, respectively for the four colors.
The peak-to-peak variations are 0.058, 0.034, 0.020, and 0.017 mag in the $u-g$, $g-r$, $r-i$, and $i-z$ colors, respectively.
The offsets in all colors display similar patterns and the variations are well correlated. 
Most offsets are negative in the range $-30 <$ RA $< 10\degr$ and  positive out of the range.
The results are consistent with an independent test of I07 based on the colors of 19,377 
red galaxies\footnote{Red galaxies serve as good color standards due to their tight color-redshift relation (Eisenstein et al. 2001), 
with a scatter of 0.12\,mag in the $u-g$ color, 0.05\,mag in the $g-r$ color, 
and 0.03\,mag in the $r-i$ and $i-z$ colors (I07).}, where they find that 
the colors are redder around RA $= -10\degr$ than around RA $= 40\degr$. 
To compare our results and those of I07 based on the galaxy colors, 
we have overplotted the latter in Fig.\,4 as blue lines, and found 
that they agree remarkably well. 
Note that I07 use a coarser spatial resolution,
i.e., a total number of 11 bins in $u-g$ and 55 bins in other colors.

\begin{figure*}
\includegraphics[width=180mm]{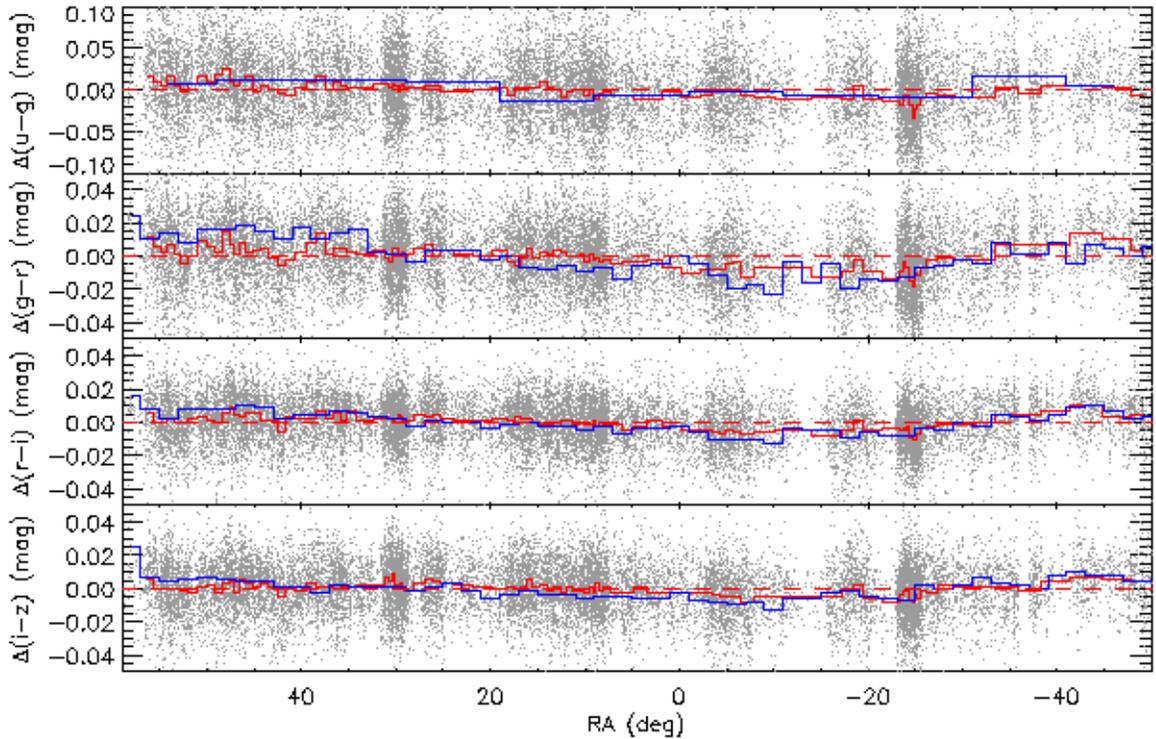}
\caption{Offsets in the $u-g$, $g-r$, $r-i$, and $i-z$ colors as a function of RA. 
The dashed lines indicate zero offsets. The sample is divided into 
100 bins in RA, with each bin containing an equal number of stars. The red solid lines denote the median offsets of each bin. 
The blue lines show the independent test results of I07 based on the colors of 19,377 red galaxies. 
Note that I07 use a coarser binning, i.e., 11 bins in $u-g$ and 55 bins in other colors.   
}
\label{}
\end{figure*}

\subsection{Spatial variations of color offsets in Dec $\delta_c^{ff}({\rm Dec})$}

After corrected for the spatial variations of color offsets in RA, 
the spatial variations of color offsets in Dec are plotted in Fig.\,5.
Following I07, the median color offsets are computed for 0.01$^\circ$􏰀
wide bins, and then smoothed using a triangular filter [i.e., $y_i$ is replaced by $0.25*(y_{i-1}+2y_i+y_{i+1})$].
The results are overplotted in Fig.\,5 and listed in Table\,3 of the Appendix.
Each bin contains about 80 stars. After the smoothing, the errors of the median color offsets of the individual bins are about 
3.0, 1.5, 1.0 and 1.0 mmag in the $u-g$, $g-r$, $r-i$, and $i-z$ colors, respectively.
The corresponding peak-to-peak variations in the four colors 
are 0.075, 0.033, 0.017, and 0.023 mag, respectively.
The offsets in all colors display similar patterns of variations and are well correlated.
For comparison, we have also overplotted in Fig.\,5 the independent test results of I07 for color $i - z$ 
based on the colors of red galaxies.
Our results are consistent with those of I07.
The test results of I07 for other colors based on the colors of red galaxies are not publicly available.  

To better illustrate the spatial structure in the color offsets, 
the two-dimensional spatial variations of offsets in the $u-g$, $g-r$, $r-i$, and $i-z$ colors are displayed in Fig.\,6.

\begin{figure*}
\includegraphics[width=180mm]{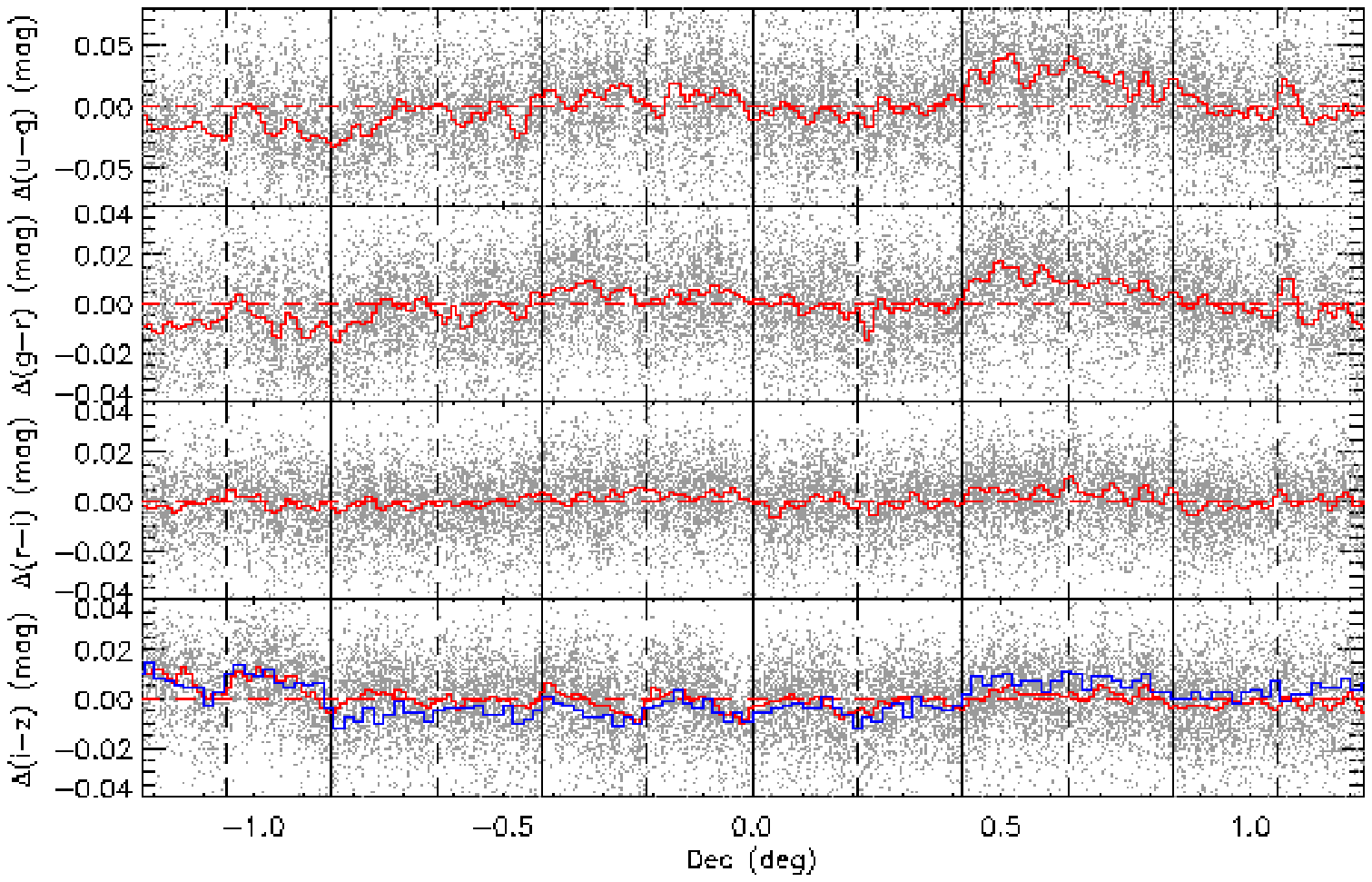}
\caption{
Offsets in the $u-g$, $g-r$, $r-i$, and $i-z$ colors as a function of Dec after corrected for $\delta_c^{ext}({\rm RA})$. 
The dashed lines indicate zero offsets. The sample is divided into
246 bins in Dec of step 0.01\degr. The red lines denote the median offsets of each bin.
The blue line denotes the test results of I07 based on the colors of red galaxies.
The vertical solid and dashed lines mark respectively the approximate boundaries between the different camera CCD columns and 
those between the two strips of the Stripe.
}
\label{}
\end{figure*}

\begin{figure}
\includegraphics[width=90mm]{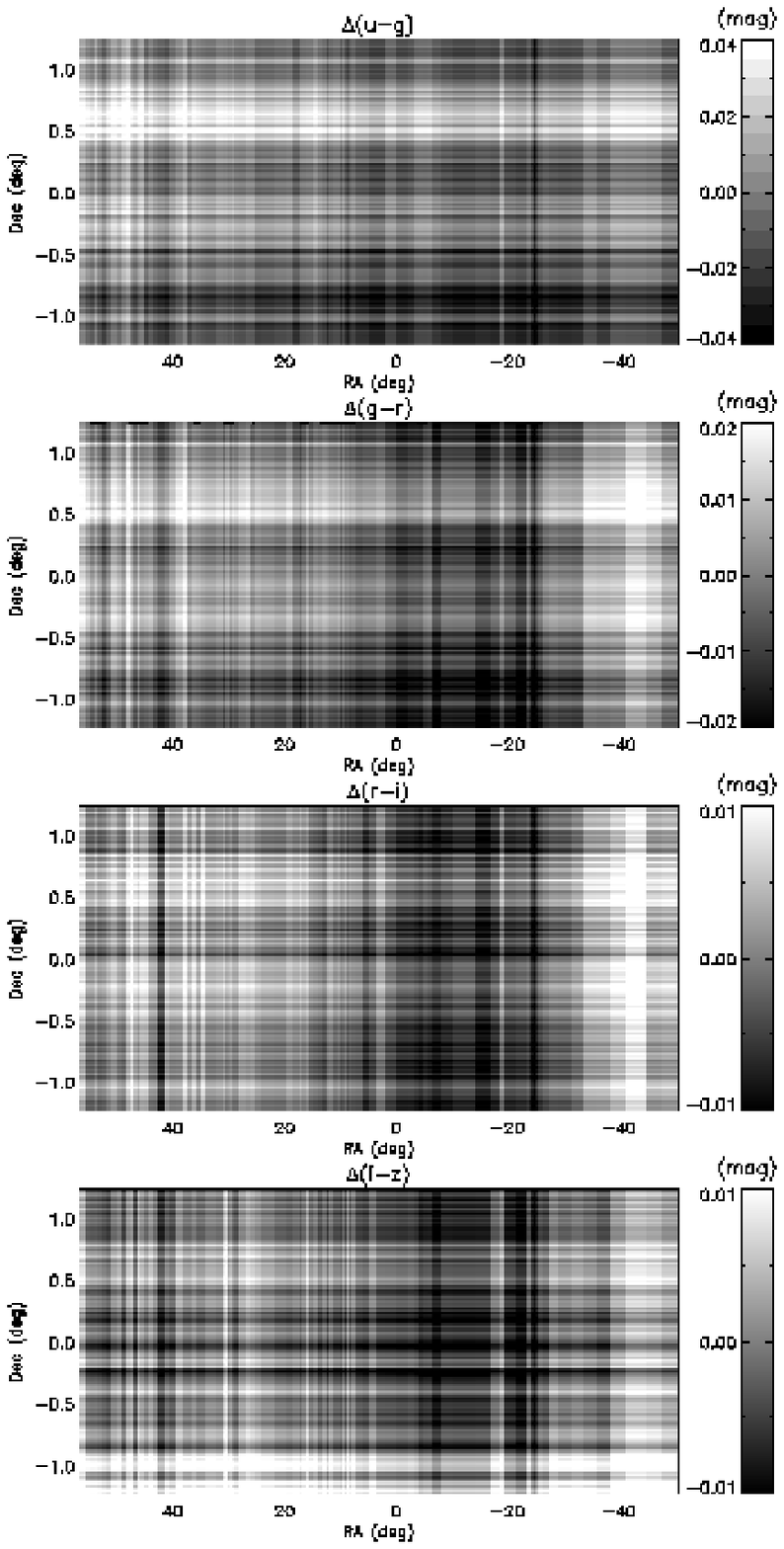}
\caption{
Spatial variations of offsets in the $u-g$, $g-r$, $r-i$, and $i-z$ colors.
A colorbar is over-plotted to the right of each panel.
}
\label{}
\end{figure}

\subsection{Dependence on magnitudes and colors}
After correcting for $\delta_c({\rm RA,Dec})$ and $\delta_c({\rm Dec})$, we investigate the possible dependence of the 
remaining color offsets on source magnitudes and colors. 
The offsets may vary with source magnitudes as a consequence of (improper corrections of) the non-linearity of the detectors,
whereas color-dependent variations may result from (improper corrections of) the bandpass variations 
amongst the different camera columns or from the 
wavelength-dependent variations of the Earth atmospheric transmissivity. 
The non-linearity of detectors have been corrected for by the SDSS photometric pipeline to an accuracy of better than 5 mmag, 
so no more corrections are applied in I07. If any effects of the non-linearity remain, they should exist in all the SDSS photometric data. 
The bandpass differences amongst the six CCD columns can lead to discrepancies in photometric colors at 1 per cent level, 
and have been corrected for by incorporating some by color terms in I07 (see their Table\,1). 
The potential wavelength-dependent variations of the Earth atmospheric transmissivity have been neglected in I07 
due to the lack of usable data. The effects are however believed to be small.
 
Fig.\,7 plots variations of $u-g$ color offsets as a function of $g$ magnitude, 
$g-r$ color offsets as a function of $r$ magnitude, $r-i$ color offsets as a function of $i$ magnitude, and 
$i-z$ color offsets as a function of $z$ magnitude for the six SDSS CCD columns.
No obvious variations with magnitudes are found except for the $i-z$ colors measured by four of the six CCD columns.
For the latter, variations over 20 mmag are observed. Since no variations in the $r-i$ color offsets are detected,
the variations of $i-z$ color offsets with $z$ magnitude are likely caused by 
(improper corrections of) the non-linearity of detectors for the $z$ band.
We have performed a 2nd-order polynomial fit to the observed variations of $i-z$ color offsets as a function of $z$ 
magnitude for the four CCD columns where the variations are seen. The results 
are overplotted in red in Fig.\,7 and the fit coefficients are listed in Table\,1.

Fig.\,8 plots color offsets in the $u-g$, $g-r$, $r-i$, and $i-z$ colors 
as a function of $g-i$  for the six SDSS CCD columns.
No obvious variations are found for stars bluer than $g-i = 1.0$\,mag. 
For redder stars, Fig.\,8 seems to indicate that there might be some small systematic variations at the 
level of about 1 mmag in $u-g$ and about 0.5 mmag in the other three colors.
The variations might be caused by the fact that red dwarfs are all local objects and, 
for at least some of them, the reddening corrections have been overestimated.
Given the very small magnitudes of variation, we have opted to not correct for those possible variations.
  
\begin{table} 
\centering
\caption{Fit coefficients for the $i-z$ color offsets as a function of $z$ magnitude$^a$. }
\label{}
\begin{tabular}{lcccc} \hline\hline
CCDs & a$_0$ & a$_1$ & a$_2$  \\
camcol = 6, Dec $= 1.05\degr$ &  1.132 & $-$0.1337 &  0.00393 \\
camcol = 5, Dec $= 0.63\degr$ &  &   &   \\
camcol = 4, Dec $= 0.21\degr$ &  &   &   \\
camcol = 3, Dec $=-0.21\degr$ &  0.290 & $-$0.0322 &  0.00088 \\
camcol = 2, Dec $=-0.63\degr$ &  0.443 & $-$0.0488 &  0.00132 \\
camcol = 1, Dec $=-1.05\degr$ &  0.496 & $-$0.0585 &  0.00172 \\
\hline
\end{tabular}
\begin{description}
\item[$^a$]  $f(z)=a_0+a_1 \times z+a_2 \times z^2$, where $z$ is the observed $z$ band magnitude. 
\end{description}
\end{table}

\begin{figure*}
\includegraphics[width=180mm]{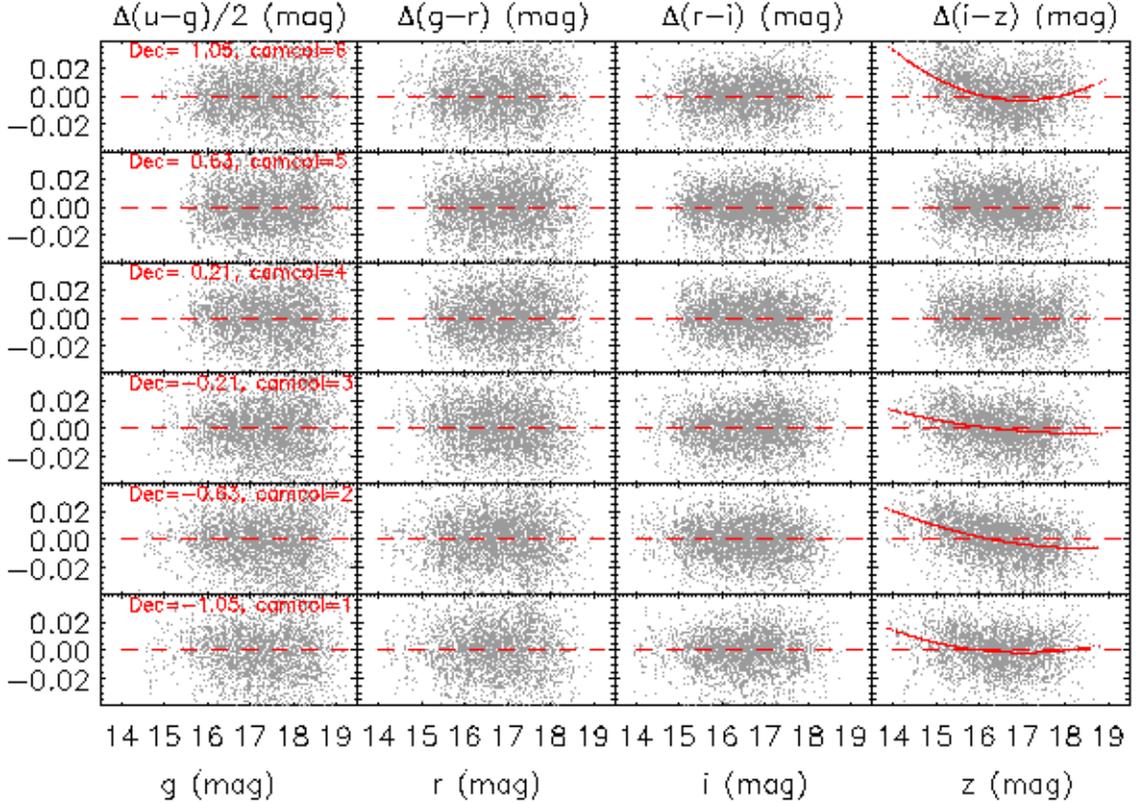}
\caption{From left to right, variations of $u-g$ color offsets as a function of $g$ magnitude,
$g-r$ color offsets as a function of $r$ magnitude, $r-i$ color offsets as a function of $i$ magnitude, and
$i-z$ color offsets as a function of $z$  magnitude for the six SDSS CCD columns, after correcting for $\delta_c^{ext}({\rm RA})$
and $\delta_c^{ff}({\rm Dec})$. Note the $u-g$ color offsets have been scaled by a factor of two.
}
\label{}
\end{figure*}

\begin{figure*}
\includegraphics[width=180mm]{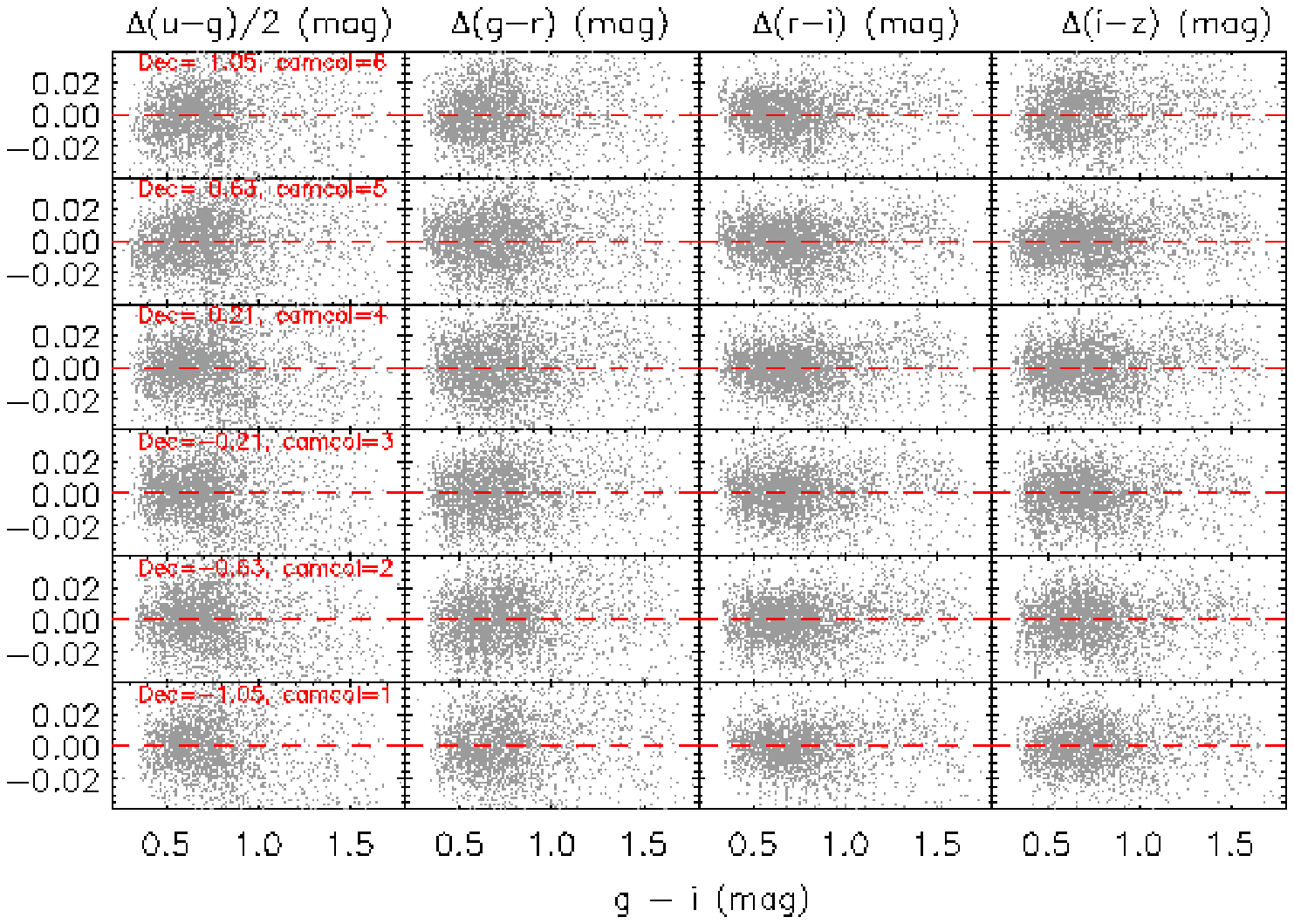}
\caption{From left to right, color offsets in the $u-g$, $g-r$, $r-i$, and $i-z$ colors
plotted against $g-i$ for the six SDSS CCD columns, after correcting for $\delta_c^{ext}({\rm RA})$ 
and $\delta_c^{ff}({\rm Dec})$.  Note the $u-g$ color offsets have been scaled by a factor of two.
}
\label{}
\end{figure*}

\subsection{Final accuracies}
There are 100 bins in $\delta_c^{ext}({\rm RA})$  and 246 bins in $\delta_c^{ff}({\rm Dec})$.
Combining together we have constructed 2D maps of $\delta_c$ of the individual colors for Stripe 82. 
Their histogram distributions are shown in Fig.\,9. 
The results suggest that I07 have achieved a color calibration accuracy better than 18, 9.9, 5.5, and 5.2 mmag,
in the $u-g$, $g-r$, $r-i$, and $i-z$ colors, respectively, and corroborate the conclusions of I07
that the rms spatial variations of the zero points in the SDSS Stripe 82 catalog must be below 0.01 mag in the $g, r$ and $i$ 
bands and very unlikely exceeding 0.02 mag in the $u$ band. 

\begin{figure}
\includegraphics[width=90mm]{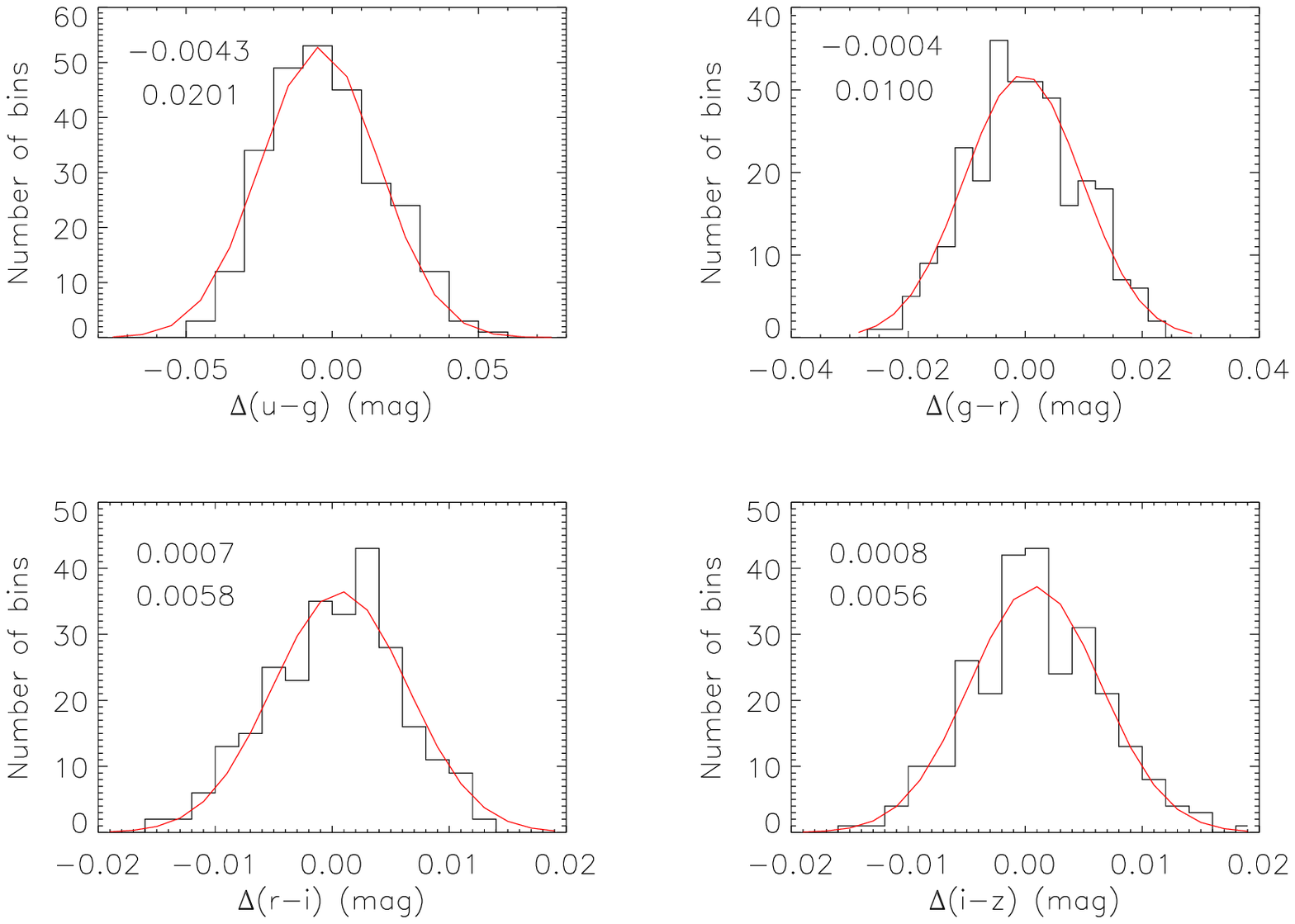}
\caption{Histogram distributions of the color offsets for the Stripe 82 region.
In each panel the red line indicates the Gaussian fitting result. The centre and sigma values are labelled.}
\label{}
\end{figure}

To further examine the color offsets, we have plotted
$\delta_c^{ext}({\rm RA})$ versus RA, $\delta_c^{ext}({\rm RA})$ versus $\delta_{g-r}^{ext}({\rm RA})$,
$\delta_c^{ff}({\rm Dec})$ versus Dec and $\delta_c^{ff}({\rm Dec})$ versus $\delta_{g-r}^{ff}({\rm Dec})$ in Fig.\,10.
Tight correlations are found between the offsets of individual colors.
The correlations are related to some problems intrinsic of the stellar locus method, 
which constrains only the shifts in combinations of two colors, the so-called 
principle colors ($s,w,x,y$; cf. Ivezi{\'c} et al. 2004), but not those in individual colors.
As a result, there are some degeneracies amongst the offsets of individual colors.
In fact, from the definitions of principle colors,
\begin{equation} 
s=-0.249\times(u-g) + 0.555\times(g-r) + 0.234,
\end{equation}
\begin{equation} 
w=-0.227\times(g-r) + 0.567\times(r-i) - 0.002\times r + 0.05,
\end{equation}
\begin{equation} 
x=0.707\times(g-r) - 0.988,
\end{equation}
\begin{equation} 
y=-0.270\times(r-i) + 0.534\times(i-z) - 0.004\times i + 0.054,
\end{equation}
we expect that 
\begin{equation} 
\delta_{u-g} = 2.23~\times~\delta_{g-r},
\end{equation}
\begin{equation} 
\delta_{r-i} = 0.40~\times~\delta_{g-r},
\end{equation}
\begin{equation} 
\delta_{i-z} = 0.50~\times~\delta_{r-i},
\end{equation}
where $2.23=0.555/0.249$, $0.40=0.227/0.567$, and $0.50=0.270/0.534$.
Note that in the above theoretical derivations, we ignore the $-0.002\times r$ term in $w$ and $-0.004\times i$ term in $y$, respectively.
We have plotted the expected relations in the right panels of Fig.\,10. 
The data follow the expected relations very well. 
There is some indication that $\delta_{u-g}^{ext}({\rm RA})$ 
may deviate slightly from the expected relation. However, the deviations are mostly caused by a few outliers 
from low-sampling regions that thus may have suffered large errors.
Note that some abnormal variations of $\delta_{i-z}^{ff}({\rm Dec})$ 
are also seen at Dec\,$< -0.84$\,$\degr$. The cause is unclear.
The above results suggest that most of the remaining calibration errors in I07 are caused by the 
degeneracies of offsets in different colors that are intrinsic of the stellar locus method.
The residuals relatively to the expected relations are estimated and marked in Fig.\,10.
Clearly, the SCR method proposed in the current work has been able to break the degeneracies and those 
residuals serve as a good indicator of the calibration accuracies of the current work.
The residuals are about 7.5, 2.5, and 3.0 mmag for the $\delta_{u-g}$ versus $\delta_{g-r}$, $\delta_{r-i}$ versus $\delta_{g-r}$ 
and $\delta_{i-z}$ versus $\delta_{g-r}$ relations, respectively, suggesting that the current calibration 
based on the SCR method should have achieved an accuracy about a few mmag.

\begin{figure*}
\includegraphics[width=180mm]{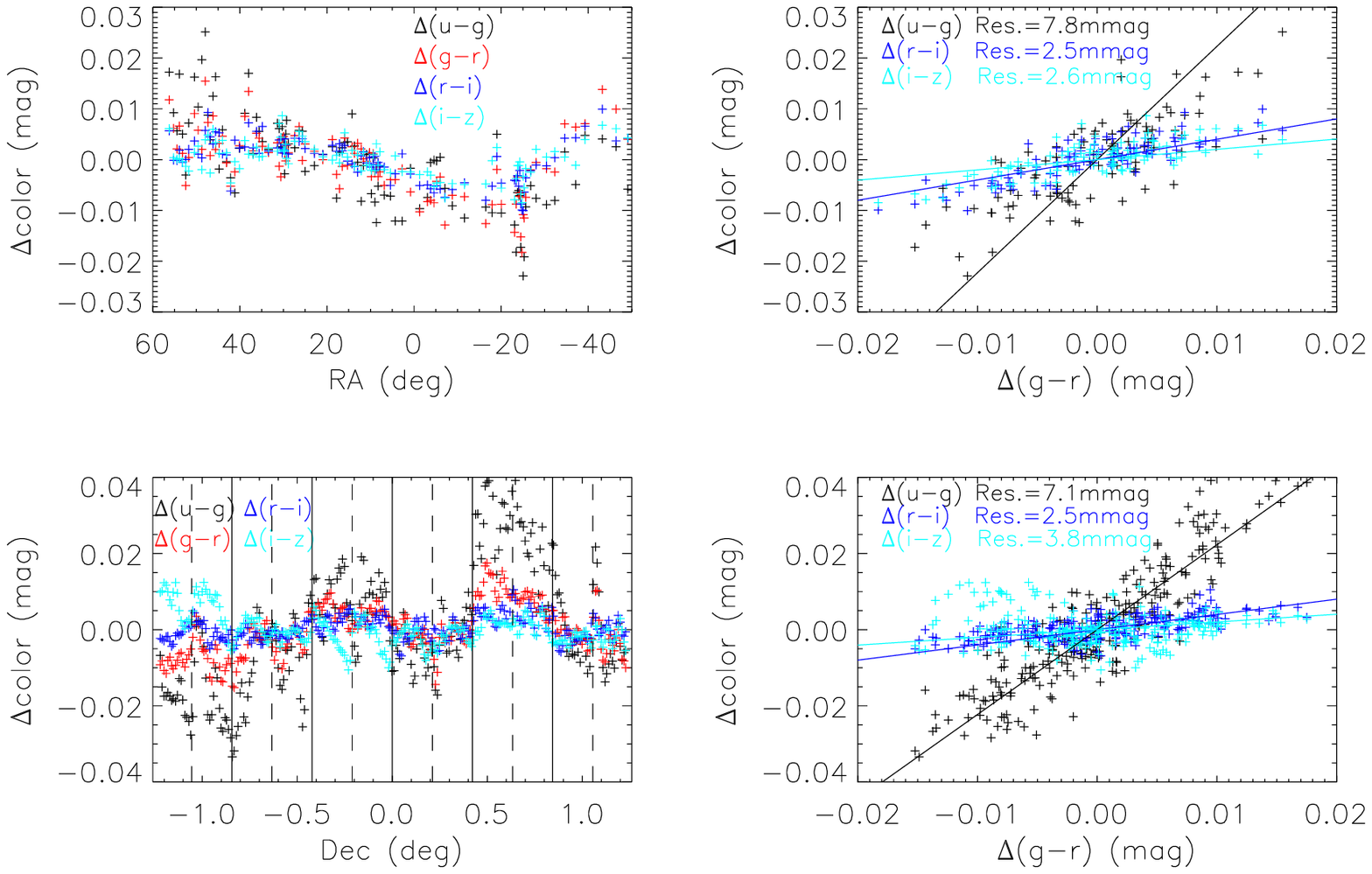}
\caption{
$\delta_c^{ext}({\rm RA})$ plotted against RA (top left) and $\delta_{g-r}^{ext}({\rm RA})$ (top right);
and $\delta_c^{ff}({\rm Dec})$ plotted against Dec (bottom left) and $\delta_{g-r}^{ff}({\rm Dec})$ (bottom right).
The lines in the right panels delineate the expected relations. The residuals relative to the expected relations are also marked.
}
\label{}
\end{figure*}

Fig.\,11 plots $\delta_c^{ext}({\rm RA})$ against the median values of \ebv~of the individual RA bins.
No dependence on \ebv~ is detected, suggesting that the calibration errors due to the 
uncertainties of reddening corrections, if any, must be very small. 

\begin{figure}
\includegraphics[width=90mm]{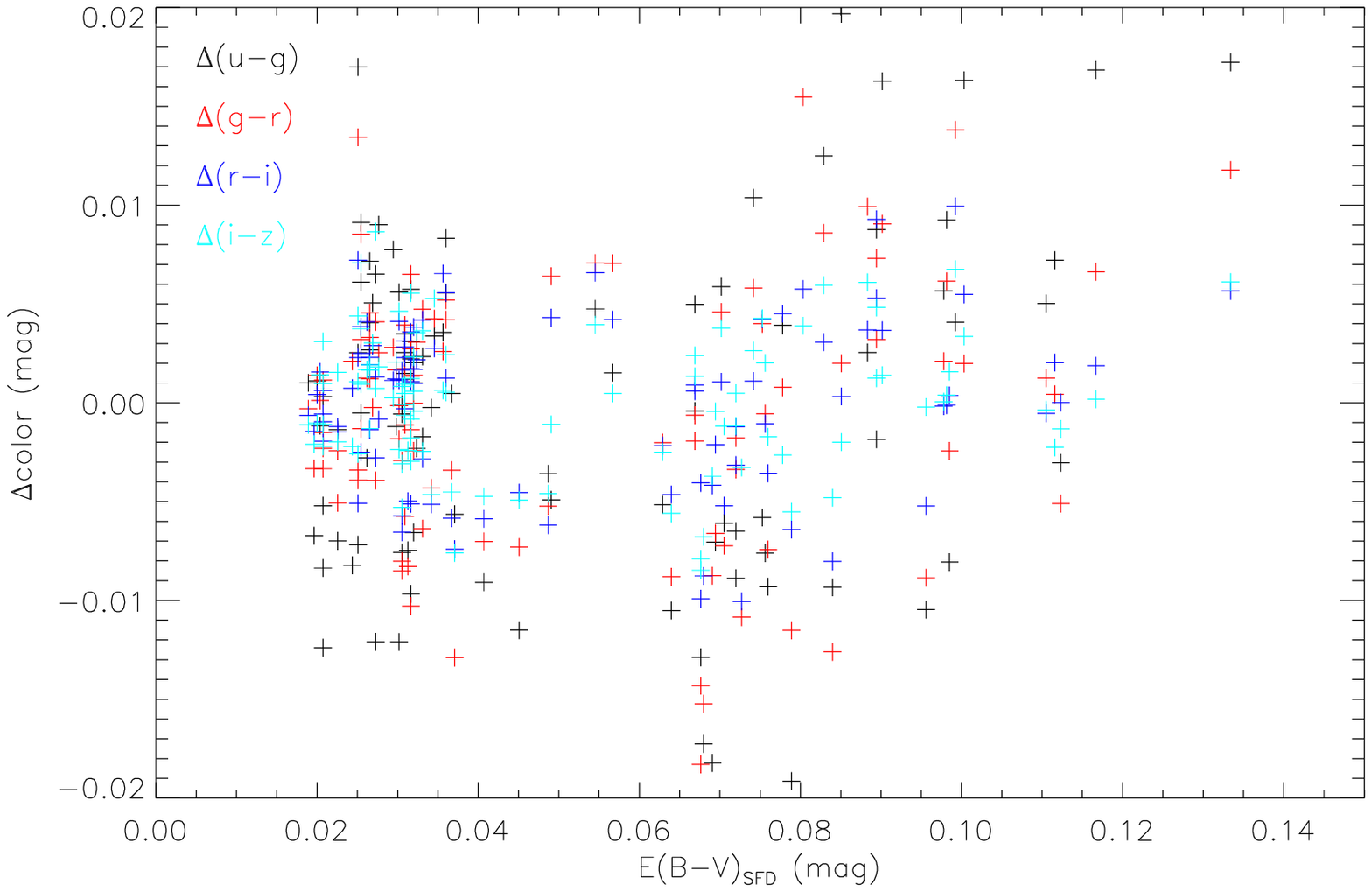}
\caption{ $\delta_c^{ext}({\rm RA})$ plotted against \ebv.
}
\label{}
\end{figure}

To further examine the calibration accuracies achieved with the SCR method, we have made a detailed comparison of our results with 
those from a completely independent approach based on the colors of red galaxies as given in I07. 
The first column of Fig.\,12 plots the differences of $\delta_c^{ext}({\rm RA})$ as given by the two approaches 
as a function of \ebv. Linear interpolations in \ebv~are used when comparing data. 
Modest correlations are found due to the different reddening coefficients used, as one would expect.
Linear regressions are carried out in order to correct for the correlations. The results are overplotted. 
The second column plots the differences against RA.
The black and red dots denote the  differences before and after corrected for the dependence of the differences on \ebv.
The large scale pattern of variations of the differences of $\delta_{u-g}^{ext}({\rm RA})$ as a function of RA is 
caused by the fact that the test results of I07 only have 11 bins in RA.
The third and fourth columns of Fig.\,12 show histogram distributions of the differences before and after the corrections, respectively.
The distributions are fitted with Gaussians whose centre and sigma values are marked.
The dispersions of the differences after the corrections 
are 8.2, 7.4, 3.0 and 2.6 mmag in the $u-g$, $g-r$, $r-i$, and $i-z$ colors, respectively.
Note that the random errors of the offsets in the $u-g$, $g-r$, $r-i$, and $i-z$ colors
are estimated at 2.8, 1.3, 0.9, and 0.9 mmag, respectively for this work,
and 6 -- 10, 3 -- 5, 1.5 -- 2.0, and 2 -- 3 mmag, respectively in I07.
Taking the random errors into account as well as the low RA resolution used in I07, 
the small dispersions suggest that we have probably achieved 
a color calibration accuracy of about or better than 5 mmag in $u-g$,  3 mmag in $g-r$, and 2 mmag in $r-i$ and $i-z$, respectively. 
Those are only a factor of two times larger than those expected from the random errors of the deduced color offsets 
alone and about two times smaller than achieved in I07.
The small dispersions also suggest that red galaxies of well determined redshifts from the modern large scale spectroscopic surveys
can also serve as excellent color standards to perform accurate color calibration in the future.

\begin{figure*}
\includegraphics[width=180mm]{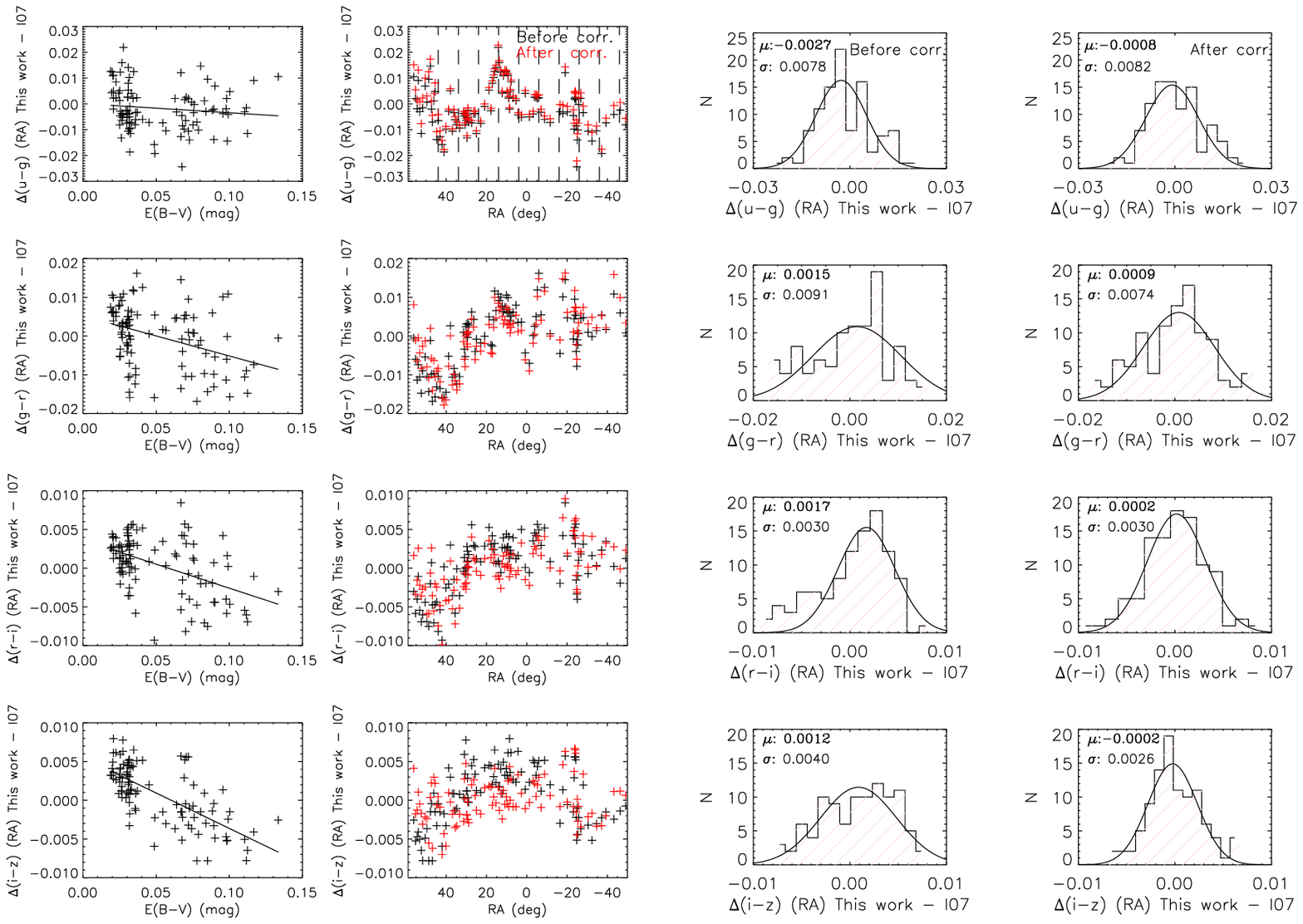}
\caption{Comparisons of values of $\delta_c^{ext}({\rm RA})$ as yielded by the SCR method proposed 
in the current work and those deduced based on the colors of red galaxies as given by I07.
The first column plots the differences of $\delta_c^{ext}({\rm RA})$ as yielded by the two approaches 
as a function of \ebv. Linear fits to the data are overplotted. 
The second column plots the differences as a function of RA before (black dots) and after (red dots) 
corrected for the dependence of the differences on \ebv. 
The dotted lines in the top panel of the second column delineate RA bins of I07.
The third column plots histogram distributions of the differences.
The lines are Gaussian fits to the distributions. The centre and sigma values are labelled.
The forth column plots the same distributions after corrected for dependence of the differences on \ebv.
}
\label{}
\end{figure*}

The top panel of Fig.\,13 plots differences of $\delta_{i-z}^{ff}({\rm Dec})$
as yielded by the SCR method and those of I07 based on the colors of red galaxies, as a function of Dec. 
The data are divided into three regimes of Dec, corresponding to the different CCD columns that exhibit different 
systematics. The variations seen in the plot are unexpected and their causes are unclear.
Histogram distributions of the differences in the three Dec regimes are shown in the
bottom panels of Fig.\,13. The dispersions vary between 2.1 -- 3.8 mmag, consistent with
the calibration accuracy achieved in the $i-z$ color.

\begin{figure}
\includegraphics[width=90mm]{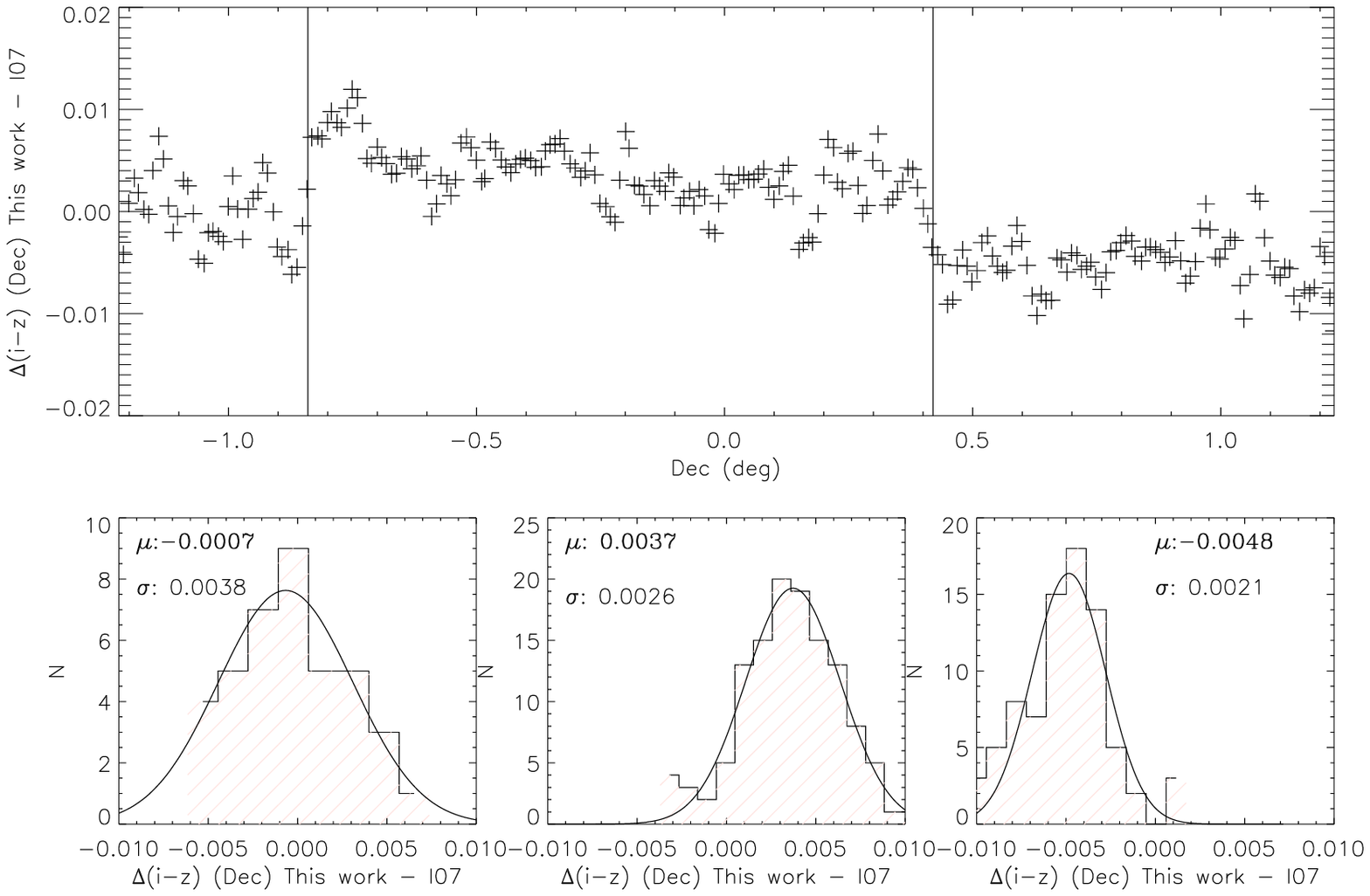}
\caption{ 
Comparison of values of $\delta_{i-z}^{ff}({\rm Dec})$ yielded by the SCR method of the current work and those 
of I07 based on the colors of red galaxies.
The data are divided into three regimes of Dec as marked by the two vertical lines.
The bottom panels plot histogram distributions of the differences in the three regimes. 
The lines are the Gaussian fits to the distributions.
The centre and sigma values are labelled.
}
\label{}
\end{figure}

To demonstrate the improvements achieved with the SCR calibration, 
we have compared the fit residuals as marked in Fig.\,3 with those after the recalibration. 
The fit residuals represent the errors of values of reddening determined with the star pair technique.
The comparisons are shown in Fig.\,14.
It is found that the residuals decrease from 39.0, 19.4, 13.3, and 14.3 mmag before the recalibration 
to 32.6, 16.1, 11.8, and 12.8 mmag after the recalibration, in the $u-g$, $g-r$, $r-i$ and $i-z$ colors, respectively, 
representing improvements of 21.4, 10.8, 6.1 and 6.4\,mmag
in the four colors, respectively.
The improvements are estimated by quadrature subtraction of the above two sets of residuals.
All the improvements are due to the decrements in color calibration errors.
Those numbers are well consistent with the color calibration accuracies before and after the recalibration. 

\begin{figure*}
\includegraphics[width=160mm]{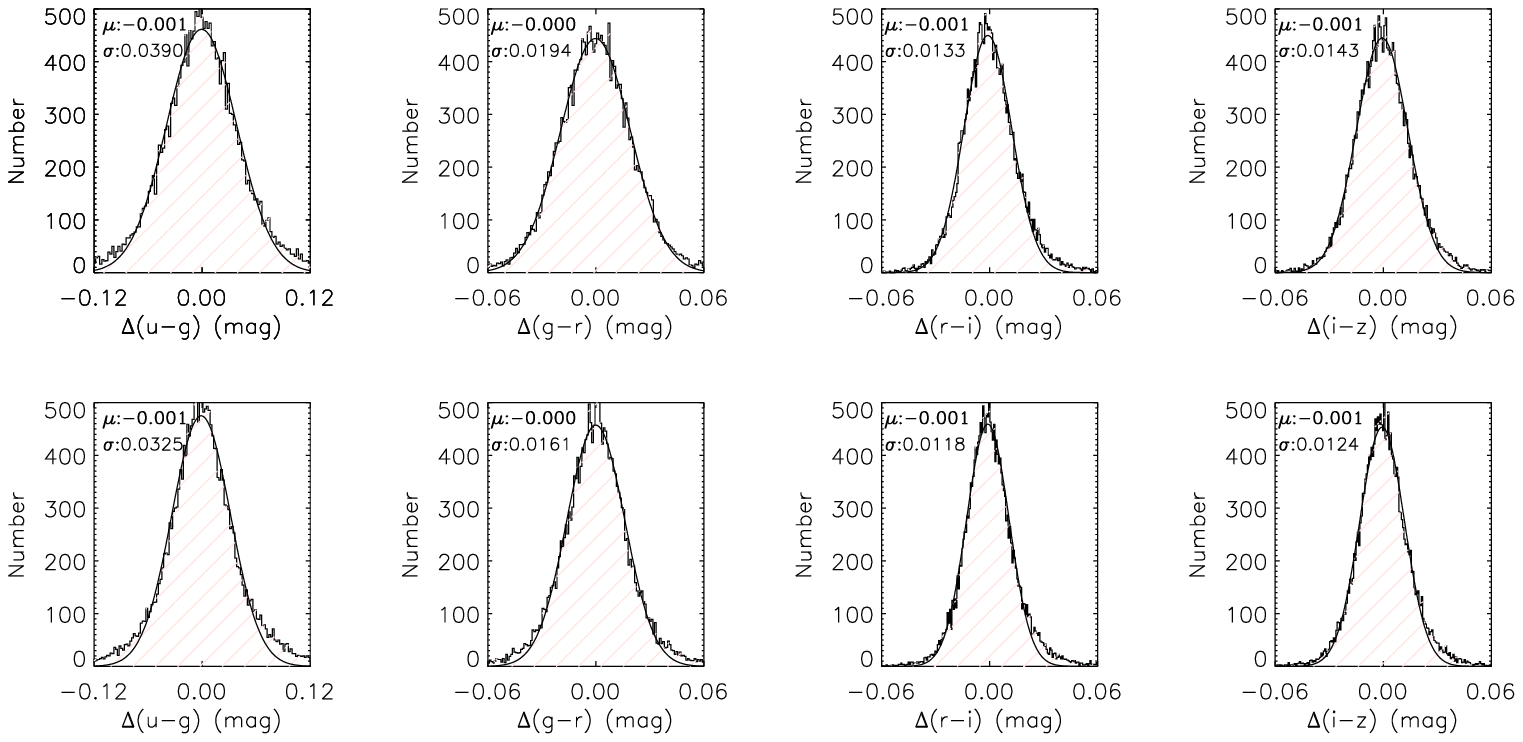}
\caption{Distributions of fit residuals before (top panels) and after (bottom panels) the recalibration. 
Also overplotted are Gaussian fits to the distributions. 
The centre and sigma values of the Gaussians are labelled.
}
\label{}
\end{figure*}

\section{Conclusions and discussions}
In this paper, we have proposed and demonstrated a spectroscopy based Stellar Color Regression 
method to perform accurate color calibration for modern imaging photometric surveys. 
The method is proposed taking advantage that we have entered an era of millions of stellar spectra 
with the SDSS, LAMOST and other on-going large scale spectroscopic surveys. 
Imaging data have been used to calibrate spectra for a long time. 
Spectroscopic data now start to benifit imaging data (see also Schlafly et al. 2011), 
by calibrating the latter to an unprecedented precision.

The SCR method requires that: 
1) There are a few well-calibrated defining fields from which the spectroscopic reference stars are selected in order
to determine the stellar intrinsic colors for a given set of stellar parameters;
2) The reddening law does not vary in a given (target) field to be calibrated; 
and 3) The values of reddening of stars spectroscopically targeted in the field are known.
With those inputs, the SCR method can be used to determine the color zero point 
of the target field by performing a linear regression between the reddening values and
the color offsets of those spectroscopically targeted stars in the field 
relative to the intrinsic values. In this way, the method brings the whole survey onto a 
uniform color scale. The method yields the reddening coefficients of the field simultaneously as a by-product.
Compared to the SLR method, the SCR method is straightforward, model-free and can be applied to regions
of different environments (Galactic halo or disk, low or high extinction) that are effectively covered by spectroscopic surveys. 
More importantly, given the facts that the stellar densities sampled by modern spectroscopic surveys such as 
the SDSS and LAMOST are typically about a few hundred per deg$^2$, 
and a precision of 50 -- 100\,K, 0.1 -- 0.25\,dex and 0.05 -- 0.1\,dex in atmospheric parameters \teff, \logg~and \feh~
has been routinely delivered by the modern stellar parameter pipelines, 
the SCR method is capable of achieving a color calibration accuracy of a few mmag.

We have applied the method to the SDSS Stripe 82 catalog carefully calibrated by I07, 
effectively the defining catalog of the SDSS photometric system with photometry internally consistent at the 1 per cent level. 
Using a total number of 23,759 stars spectroscopically targeted within the Stripe, we have mapped the small yet significant 
spatial variations of color calibration errors present in the catalog, achieving a remarkable improvement by a factor of 2 -- 3.
The calibration errors of different colors are strongly correlated, and arise 
from the degeneracies intrinsic to the stellar locus method used by I07.
Possible dependence of the errors on color and magnitude is also examined.
Small but significant magnitude dependent errors are found for $z$-band for some CCDs. 
Such errors should be present in all SDSS photometric data. 
Our results are compared with those of I07 from a completely independent test based on the colors of red galaxies.
The comparisons show that the SCR method have achieved
a color calibration accuracy of about or better than 5 mmag in $u-g$,  3 mmag in $g-r$, and 2 mmag in $r-i$ and $i-z$, respectively. 
The achieved uncertainties are only a factor of two times larger than those expected the random errors of color offsets. 
The accuracies are consistent with the small residuals of the 
calibration errors of the individual colors in I07 relative to the expected relations (cf. right panels in Fig.\,10)
and the improvements in determinations of reddening with the star pair technique.

Given the power of the SCR method, applying the method to the existing imaging photometric 
data within the footprints of the existing SDSS and LAMOST surveys 
will significantly improve the color calibration accuracies of those photometric surveys, 
such as the SDSS, the Pan-STARRS1, the Xuyi Schmidt Telescope Photometric Survey of the Galactic anti-centre (XSTPS-GAC; Liu et al. 2014) and
the Isaac Newton Telescope Photometric H$\alpha$ Survey (IPHAS; Drew et al. 2005). 
The method also provides a promising way to calibrate future large scale imaging surveys such as the LSST if there 
will also be new accompanying spectroscopic surveys in the southern hemisphere. 

With colors calibrated to a few or even one mmag accuracy for the modern imaging surveys, it will greatly advance the fields of  
object classifications, studies of the large scale structures with the photometric redshifts 
and studies of the Galactic topographical structures.
For example,  Yuan et al. (2014b) provide metallicity dependent stellar loci 
in terms of the SDSS colors using the Stripe 82 photometric data
recalibrated with the SCR method in the current work
and find that the intrinsic widths of the stellar loci are at maximum of a few mmag if not zero.
As a consequence of those unprecedented accurate stellar loci, the photometric data reveal 
clear signatures of stellar binarity.  
Based on the metallicity dependent stellar loci and using a subsample of the same set of data, 
Yuan et al. (2014c, submitted) present a model-free estimate of binary fraction for field FGK stars.
Also based on the metallicity dependent stellar loci, Yuan et al. (2014d, to be submitted) develop a method 
capable of simultaneously determining reddening, metallicity, intrinsic colors, and distance for individual stars 
based on multi-band photometric data alone, thus 
opening up the possibilities of topographical studies of the Galactic disk(s)
based on large scale photometric data alone (Yuan et al. 2014e, in preparation).

The SCR method requires the reddening coefficients are constant for a given field.
For this method, there are some degeneracies between the adopted reddening coefficients and 
the resultant color zero point offsets. 
In the Galactic disk, the reddening law may vary significantly on relatively small scales. 
We will investigate and map the spatial variations of reddening law across the Galactic disk 
in a forthcoming paper utilizing the LAMOST spectroscopic data.

The SCR method works best for objects cataloged from flat-fielded images.
However, large scale smoothly varying errors in flat-fields can be detected and mapped using the method 
if there are sufficient number of spectroscopically targeted stars, as shown in the current paper.

As a closing remark, it is worth emphasizing that the red galaxies 
of well determined redshifts observed in modern large scale spectroscopic surveys  
can also serve as excellent color standards to perform independent accurate color calibration. 

\vspace{7mm} \noindent {\bf Acknowledgments}{
We would like to thank the referee for his/her valuable comments.
We also acknowledge the stimulating discussions with Mike Gladder during the 1st Kavli Astrophysics Symposium. 
This work is supported by National Key Basic Research Program of China 2014CB845700 
and China Postdoctoral Science special Foundation 2014T70011.
This work has made use of data products from the Sloan Digital Sky Survey.
}

\appendix

\section {Color offsets as a function of RA and Dec.}
\begin{table*}
\centering
\caption{Color offsets as a function of RA, $\delta_c^{ext}({\rm RA})$.}
\label{} \tiny 
\begin{tabular}{rrrrrrrrrrrrrrr} \hline\hline
 RA   & $u-g$ &  $g-r$ & $r-i$  & $i-z$  & RA   & $u-g$ &  $g-r$ & $r-i$  & $i-z$  & RA   & $u-g$ &  $g-r$ & $r-i$  & $i-z$   \\
 (deg) & (mag) &  (mag) & (mag)  & (mag)  &(deg)   & (mag) &  (mag) & (mag)  & (mag)  & (deg)   & (mag) & (mag)& (mag)  & (mag)   \\\hline
$-$49.10 & $-$0.006 &  0.004 &  0.004 &  0.004 & $-$46.39 &  0.003 &  0.010 &  0.004 &  0.006 & $-$43.25 &  0.004 &  0.014 &  0.010 &  0.007 \\
$-$39.29 &  0.005 &  0.007 &  0.007 &  0.004 & $-$37.14 & $-$0.005 &  0.006 &  0.004 & $-$0.001 & $-$34.60 &  0.002 &  0.007 &  0.004 &  0.000 \\
$-$32.33 & $-$0.008 & $-$0.002 &  0.000 &  0.002 & $-$30.43 & $-$0.009 & $-$0.002 & $-$0.001 &  0.000 & $-$28.04 & $-$0.008 & $-$0.001 & $-$0.001 &  0.002 \\
$-$26.79 & $-$0.005 & $-$0.002 & $-$0.002 & $-$0.002 & $-$25.68 & $-$0.009 & $-$0.007 & $-$0.004 & $-$0.002 & $-$25.31 & $-$0.019 & $-$0.012 & $-$0.006 & $-$0.006 \\
$-$25.05 & $-$0.023 & $-$0.011 & $-$0.010 & $-$0.003 & $-$24.81 & $-$0.033 & $-$0.018 & $-$0.010 & $-$0.008 & $-$24.53 & $-$0.017 & $-$0.015 & $-$0.009 & $-$0.007 \\
$-$24.28 & $-$0.007 & $-$0.007 & $-$0.002 & $-$0.000 & $-$24.04 & $-$0.006 & $-$0.007 & $-$0.005 & $-$0.001 & $-$23.78 & $-$0.006 & $-$0.003 & $-$0.003 & $-$0.001 \\
$-$23.48 & $-$0.018 & $-$0.009 & $-$0.004 & $-$0.004 & $-$23.10 & $-$0.013 & $-$0.014 & $-$0.004 & $-$0.008 & $-$20.06 & $-$0.011 & $-$0.009 & $-$0.005 & $-$0.006 \\
$-$19.02 &  0.005 & $-$0.002 &  0.001 &  0.002 & $-$17.91 & $-$0.010 & $-$0.009 & $-$0.005 & $-$0.000 & $-$16.47 & $-$0.009 & $-$0.013 & $-$0.008 & $-$0.005 \\
$-$11.58 & $-$0.012 & $-$0.007 & $-$0.005 & $-$0.005 &  $-$8.79 & $-$0.009 & $-$0.007 & $-$0.006 & $-$0.005 &  $-$7.09 & $-$0.006 & $-$0.013 & $-$0.007 & $-$0.008 \\
 $-$6.00 &  0.000 & $-$0.003 & $-$0.006 & $-$0.005 &  $-$5.20 & $-$0.000 & $-$0.004 & $-$0.005 & $-$0.005 &  $-$4.57 & $-$0.000 & $-$0.009 & $-$0.007 & $-$0.005 \\
 $-$3.80 & $-$0.001 & $-$0.008 & $-$0.006 & $-$0.003 &  $-$3.10 & $-$0.007 & $-$0.008 & $-$0.005 & $-$0.002 &  $-$1.29 & $-$0.010 & $-$0.010 & $-$0.005 & $-$0.003 \\
  0.81 & $-$0.002 & $-$0.006 & $-$0.003 & $-$0.002 &   2.74 & $-$0.012 & $-$0.002 &  0.001 & $-$0.002 &   4.32 & $-$0.012 & $-$0.004 & $-$0.003 &  0.001 \\
  5.42 & $-$0.007 & $-$0.004 & $-$0.005 & $-$0.003 &   6.83 & $-$0.007 & $-$0.005 & $-$0.001 & $-$0.002 &   7.83 & $-$0.001 & $-$0.002 & $-$0.001 &  0.002 \\
  8.24 & $-$0.005 & $-$0.003 & $-$0.001 & $-$0.001 &   8.64 & $-$0.012 & $-$0.001 & $-$0.002 &  0.003 &   9.00 & $-$0.008 & $-$0.002 &  0.001 & $-$0.002 \\
  9.38 &  0.000 &  0.001 & $-$0.001 &  0.001 &   9.83 & $-$0.007 & $-$0.003 & $-$0.001 & $-$0.002 &  10.21 &  0.001 &  0.001 &  0.000 & $-$0.001 \\
 10.88 &  0.001 & $-$0.000 & $-$0.001 & $-$0.001 &  11.63 & $-$0.001 &  0.000 &  0.002 &  0.001 &  12.15 & $-$0.008 &  0.002 &  0.001 & $-$0.002 \\
 12.94 & $-$0.001 &  0.003 & $-$0.002 &  0.001 &  13.67 &  0.003 &  0.001 & $-$0.001 & $-$0.001 &  14.25 &  0.009 &  0.003 & $-$0.001 &  0.002 \\
 15.29 &  0.003 & $-$0.001 & $-$0.000 &  0.000 &  15.94 &  0.003 &  0.004 &  0.003 &  0.005 &  16.49 &  0.002 & $-$0.001 &  0.002 &  0.001 \\
 17.13 &  0.001 & $-$0.006 &  0.003 & $-$0.001 &  17.82 & $-$0.008 & $-$0.003 &  0.002 &  0.001 &  18.77 &  0.003 &  0.004 &  0.003 & $-$0.000 \\
 20.94 &  0.001 & $-$0.000 &  0.001 &  0.001 &  22.89 & $-$0.001 &  0.002 &  0.001 &  0.002 &  25.00 & $-$0.002 &  0.003 &  0.002 &  0.004 \\
 25.67 &  0.006 & $-$0.000 &  0.004 &  0.005 &  26.42 &  0.006 &  0.006 &  0.004 &  0.006 &  27.38 &  0.002 &  0.005 &  0.004 &  0.004 \\
 28.87 &  0.003 &  0.001 &  0.001 & $-$0.001 &  29.26 &  0.008 &  0.003 &  0.001 &  0.000 &  29.48 &  0.005 & $-$0.000 &  0.003 &  0.003 \\
 29.73 & $-$0.003 &  0.001 &  0.004 &  0.002 &  29.97 &  0.004 &  0.005 &  0.002 &  0.002 &  30.30 &  0.007 &  0.004 &  0.001 &  0.009 \\
 30.60 &  0.003 & $-$0.003 &  0.002 &  0.004 &  30.98 &  0.006 & $-$0.001 &  0.004 &  0.007 &  31.37 &  0.007 &  0.003 &  0.002 &  0.002 \\
 33.40 &  0.002 &  0.001 &  0.002 &  0.002 &  34.83 &  0.004 &  0.003 &  0.007 &  0.001 &  35.59 &  0.006 &  0.004 &  0.001 &  0.001 \\
 36.18 &  0.008 &  0.005 &  0.006 &  0.002 &  36.95 &  0.001 &  0.002 &  0.002 &  0.001 &  37.99 &  0.017 &  0.013 &  0.007 &  0.001 \\
 38.86 &  0.009 &  0.009 &  0.003 &  0.004 &  40.12 &  0.002 &  0.003 &  0.002 & $-$0.002 &  41.16 & $-$0.007 & $-$0.002 &  0.004 & $-$0.000 \\
 42.13 & $-$0.004 & $-$0.005 & $-$0.006 & $-$0.005 &  43.09 & $-$0.000 & $-$0.001 &  0.001 &  0.001 &  44.01 &  0.006 &  0.005 &  0.001 &  0.004 \\
 44.76 & $-$0.002 &  0.003 &  0.005 &  0.001 &  45.57 &  0.016 &  0.002 &  0.005 &  0.003 &  46.19 &  0.012 &  0.009 &  0.003 &  0.006 \\
 46.83 &  0.004 &  0.001 &  0.005 & $-$0.003 &  47.34 &  0.009 &  0.007 &  0.009 &  0.005 &  47.95 &  0.025 &  0.015 &  0.006 &  0.004 \\
 48.74 &  0.020 &  0.002 &  0.000 & $-$0.002 &  49.58 &  0.010 &  0.006 &  0.001 &  0.003 &  50.44 &  0.016 &  0.009 &  0.004 &  0.001 \\
 51.75 &  0.007 &  0.000 &  0.002 & $-$0.002 &  52.42 & $-$0.003 & $-$0.005 &  0.000 & $-$0.001 &  53.18 &  0.005 &  0.001 & $-$0.001 & $-$0.000 \\
 53.91 &  0.017 &  0.007 &  0.002 &  0.000 &  54.46 &  0.006 &  0.002 & $-$0.000 &  0.000 &  55.21 &  0.009 &  0.006 & $-$0.000 &  0.000 \\
 56.23 &  0.017 &  0.012 &  0.006 &  0.006 & & & & & & & & & & \\
\hline
\end{tabular}
\end{table*}

\begin{table*}
\centering
\caption{Color offsets as a function of Dec, $\delta_c^{ff}({\rm Dec})$, after corrected for $\delta_c^{ext}({\rm RA})$.}
\label{} \tiny 
\begin{tabular}{rrrrrrrrrrrrrrr} \hline\hline
 Dec   & $u-g$ &  $g-r$ & $r-i$  & $i-z$  & Dec   & $u-g$ &  $g-r$ & $r-i$  & $i-z$  & Dec   & $u-g$ &  $g-r$ & $r-i$  & $i-z$   \\
 (deg) & (mag) &  (mag) & (mag)  & (mag)  &(deg)   & (mag) &  (mag) & (mag)  & (mag)  & (deg)   & (mag) & (mag)& (mag)  & (mag)   \\\hline
$-$1.22 & $-$0.007 & $-$0.009 & $-$0.002 &  0.010 & $-$1.21 & $-$0.015 & $-$0.008 & $-$0.003 &  0.010 & $-$1.20 & $-$0.017 & $-$0.011 & $-$0.003 &  0.012 \\
$-$1.19 & $-$0.017 & $-$0.011 & $-$0.002 &  0.012 & $-$1.18 & $-$0.017 & $-$0.008 & $-$0.002 &  0.010 & $-$1.17 & $-$0.017 & $-$0.007 & $-$0.003 &  0.008 \\
$-$1.16 & $-$0.019 & $-$0.009 & $-$0.004 &  0.006 & $-$1.15 & $-$0.018 & $-$0.010 & $-$0.004 &  0.010 & $-$1.14 & $-$0.019 & $-$0.009 & $-$0.002 &  0.012 \\
$-$1.13 & $-$0.017 & $-$0.008 & $-$0.001 &  0.010 & $-$1.12 & $-$0.013 & $-$0.006 & $-$0.001 &  0.005 & $-$1.11 & $-$0.013 & $-$0.006 & $-$0.001 &  0.002 \\
$-$1.10 & $-$0.018 & $-$0.009 & $-$0.001 &  0.001 & $-$1.09 & $-$0.022 & $-$0.010 &  0.001 &  0.000 & $-$1.08 & $-$0.023 & $-$0.008 &  0.002 &  0.002 \\
$-$1.07 & $-$0.024 & $-$0.007 &  0.000 &  0.003 & $-$1.06 & $-$0.028 & $-$0.005 &  0.001 &  0.002 & $-$1.05 & $-$0.023 & $-$0.002 &  0.005 &  0.005 \\
$-$1.04 & $-$0.008 &  0.001 &  0.004 &  0.009 & $-$1.03 &  0.001 &  0.004 &  0.002 &  0.011 & $-$1.02 &  0.002 &  0.001 &  0.002 &  0.009 \\
$-$1.01 &  0.002 & $-$0.003 &  0.001 &  0.007 & $-$1.00 & $-$0.001 & $-$0.005 &  0.002 &  0.010 & $-$0.99 & $-$0.004 & $-$0.006 &  0.002 &  0.012 \\
$-$0.98 & $-$0.013 & $-$0.005 &  0.000 &  0.011 & $-$0.97 & $-$0.022 & $-$0.008 & $-$0.003 &  0.009 & $-$0.96 & $-$0.024 & $-$0.012 & $-$0.005 &  0.009 \\
$-$0.95 & $-$0.022 & $-$0.014 & $-$0.004 &  0.008 & $-$0.94 & $-$0.017 & $-$0.009 & $-$0.002 &  0.007 & $-$0.93 & $-$0.010 & $-$0.004 &  0.000 &  0.009 \\
$-$0.92 & $-$0.013 & $-$0.005 & $-$0.001 &  0.009 & $-$0.91 & $-$0.021 & $-$0.010 & $-$0.003 &  0.006 & $-$0.90 & $-$0.026 & $-$0.013 & $-$0.002 &  0.002 \\
$-$0.89 & $-$0.027 & $-$0.014 & $-$0.002 &  0.001 & $-$0.88 & $-$0.023 & $-$0.011 & $-$0.003 &  0.002 & $-$0.87 & $-$0.024 & $-$0.009 & $-$0.003 &  0.000 \\
$-$0.86 & $-$0.026 & $-$0.008 & $-$0.001 & $-$0.004 & $-$0.85 & $-$0.028 & $-$0.009 & $-$0.000 & $-$0.005 & $-$0.84 & $-$0.033 & $-$0.015 & $-$0.003 & $-$0.005 \\
$-$0.83 & $-$0.032 & $-$0.015 & $-$0.005 & $-$0.004 & $-$0.82 & $-$0.028 & $-$0.011 & $-$0.004 & $-$0.002 & $-$0.81 & $-$0.025 & $-$0.008 & $-$0.002 & $-$0.000 \\
$-$0.80 & $-$0.017 & $-$0.007 & $-$0.001 &  0.000 & $-$0.79 & $-$0.014 & $-$0.008 & $-$0.001 &  0.000 & $-$0.78 & $-$0.023 & $-$0.008 & $-$0.001 &  0.002 \\
$-$0.77 & $-$0.024 & $-$0.007 &  0.002 &  0.004 & $-$0.76 & $-$0.018 & $-$0.004 &  0.001 &  0.003 & $-$0.75 & $-$0.014 &  0.000 & $-$0.003 &  0.002 \\
$-$0.74 & $-$0.011 &  0.002 & $-$0.004 &  0.003 & $-$0.73 & $-$0.008 &  0.002 & $-$0.002 &  0.002 & $-$0.72 & $-$0.002 & $-$0.001 & $-$0.001 &  0.001 \\
$-$0.71 &  0.006 & $-$0.002 & $-$0.000 &  0.003 & $-$0.70 &  0.004 & $-$0.001 & $-$0.002 &  0.003 & $-$0.69 & $-$0.003 & $-$0.003 & $-$0.004 &  0.000 \\
$-$0.68 & $-$0.003 & $-$0.003 & $-$0.004 & $-$0.001 & $-$0.67 & $-$0.001 &  0.002 & $-$0.002 & $-$0.002 & $-$0.66 & $-$0.001 &  0.002 & $-$0.001 & $-$0.003 \\
$-$0.65 & $-$0.002 & $-$0.000 & $-$0.000 & $-$0.003 & $-$0.64 & $-$0.000 & $-$0.003 & $-$0.001 & $-$0.002 & $-$0.63 &  0.002 & $-$0.005 & $-$0.002 & $-$0.001 \\
$-$0.62 &  0.001 & $-$0.008 & $-$0.000 &  0.000 & $-$0.61 & $-$0.003 & $-$0.006 & $-$0.000 &  0.001 & $-$0.60 & $-$0.010 & $-$0.001 & $-$0.002 & $-$0.001 \\
$-$0.59 & $-$0.015 & $-$0.005 & $-$0.002 & $-$0.005 & $-$0.58 & $-$0.016 & $-$0.010 & $-$0.000 & $-$0.003 & $-$0.57 & $-$0.013 & $-$0.009 & $-$0.001 & $-$0.001 \\
$-$0.56 & $-$0.005 & $-$0.005 & $-$0.002 & $-$0.001 & $-$0.55 & $-$0.005 & $-$0.004 & $-$0.002 & $-$0.001 & $-$0.54 & $-$0.011 & $-$0.003 & $-$0.002 & $-$0.001 \\
$-$0.53 & $-$0.005 & $-$0.000 & $-$0.001 & $-$0.000 & $-$0.52 &  0.004 &  0.001 & $-$0.000 &  0.000 & $-$0.51 &  0.003 &  0.001 &  0.001 & $-$0.001 \\
$-$0.50 & $-$0.001 & $-$0.000 &  0.001 &  0.000 & $-$0.49 & $-$0.007 & $-$0.002 & $-$0.002 & $-$0.001 & $-$0.48 & $-$0.018 & $-$0.004 & $-$0.002 & $-$0.002 \\
$-$0.47 & $-$0.026 & $-$0.006 &  0.000 & $-$0.003 & $-$0.46 & $-$0.021 & $-$0.005 &  0.002 & $-$0.003 & $-$0.45 & $-$0.007 & $-$0.001 &  0.002 & $-$0.003 \\
$-$0.44 &  0.007 &  0.005 &  0.002 & $-$0.002 & $-$0.43 &  0.009 &  0.005 &  0.003 & $-$0.001 & $-$0.42 &  0.007 &  0.003 &  0.004 &  0.003 \\
$-$0.41 &  0.013 &  0.004 &  0.003 &  0.006 & $-$0.40 &  0.014 &  0.006 &  0.001 &  0.005 & $-$0.39 &  0.007 &  0.007 & $-$0.000 &  0.003 \\
$-$0.38 &  0.004 &  0.006 & $-$0.001 &  0.002 & $-$0.37 & $-$0.002 &  0.005 & $-$0.002 &  0.001 & $-$0.36 & $-$0.003 &  0.007 &  0.001 &  0.002 \\
$-$0.35 &  0.003 &  0.008 &  0.003 &  0.002 & $-$0.34 &  0.009 &  0.008 &  0.004 &  0.001 & $-$0.33 &  0.013 &  0.010 &  0.004 & $-$0.000 \\
$-$0.32 &  0.012 &  0.010 &  0.002 & $-$0.002 & $-$0.31 &  0.007 &  0.007 &  0.000 & $-$0.003 & $-$0.30 &  0.006 &  0.003 & $-$0.000 & $-$0.003 \\
$-$0.29 &  0.013 &  0.002 &  0.002 & $-$0.003 & $-$0.28 &  0.016 &  0.004 &  0.004 & $-$0.005 & $-$0.27 &  0.015 &  0.005 &  0.003 & $-$0.005 \\
$-$0.26 &  0.019 &  0.006 &  0.002 & $-$0.006 & $-$0.25 &  0.017 &  0.006 &  0.004 & $-$0.007 & $-$0.24 &  0.007 &  0.003 &  0.004 & $-$0.008 \\
$-$0.23 &  0.004 &  0.000 &  0.005 & $-$0.011 & $-$0.22 &  0.007 & $-$0.000 &  0.006 & $-$0.008 & $-$0.21 &  0.002 &  0.000 &  0.004 & $-$0.001 \\
$-$0.20 & $-$0.004 &  0.002 &  0.004 &  0.005 & $-$0.19 & $-$0.009 &  0.002 &  0.003 &  0.003 & $-$0.18 & $-$0.007 &  0.000 &  0.001 &  0.001 \\
$-$0.17 &  0.007 &  0.002 &  0.002 &  0.002 & $-$0.16 &  0.019 &  0.006 &  0.003 &  0.002 & $-$0.15 &  0.018 &  0.004 &  0.003 &  0.004 \\
$-$0.14 &  0.008 &  0.001 &  0.003 &  0.004 & $-$0.13 &  0.004 &  0.002 &  0.001 &  0.001 & $-$0.12 &  0.010 &  0.006 &  0.000 & $-$0.001 \\
$-$0.11 &  0.013 &  0.006 &  0.001 & $-$0.000 & $-$0.10 &  0.009 &  0.004 &  0.002 &  0.001 & $-$0.09 &  0.004 &  0.005 &  0.003 &  0.000 \\
$-$0.08 &  0.008 &  0.008 &  0.004 & $-$0.001 & $-$0.07 &  0.015 &  0.007 &  0.003 & $-$0.003 & $-$0.06 &  0.011 &  0.005 &  0.002 & $-$0.006 \\
$-$0.05 &  0.009 &  0.005 &  0.004 & $-$0.007 & $-$0.04 &  0.012 &  0.005 &  0.005 & $-$0.006 & $-$0.03 &  0.010 &  0.004 &  0.004 & $-$0.008 \\
$-$0.02 &  0.003 &  0.003 &  0.001 & $-$0.010 & $-$0.01 & $-$0.003 &  0.001 &  0.000 & $-$0.009 & $-$0.00 & $-$0.011 &  0.001 & $-$0.000 & $-$0.004 \\
 0.01 & $-$0.011 &  0.002 & $-$0.001 & $-$0.003 &  0.02 & $-$0.004 &  0.002 & $-$0.000 & $-$0.003 &  0.03 & $-$0.006 & $-$0.000 & $-$0.002 & $-$0.002 \\
 0.04 & $-$0.008 & $-$0.001 & $-$0.006 & $-$0.001 &  0.05 & $-$0.001 &  0.000 & $-$0.007 & $-$0.001 &  0.06 &  0.004 &  0.003 & $-$0.003 & $-$0.000 \\
 0.07 &  0.001 &  0.004 & $-$0.001 &  0.000 &  0.08 & $-$0.002 &  0.003 & $-$0.002 &  0.001 &  0.09 & $-$0.004 & $-$0.001 & $-$0.003 &  0.000 \\
 0.10 & $-$0.009 & $-$0.005 & $-$0.002 & $-$0.002 &  0.11 & $-$0.014 & $-$0.004 &  0.000 & $-$0.002 &  0.12 & $-$0.012 & $-$0.002 &  0.002 & $-$0.000 \\
 0.13 & $-$0.005 & $-$0.001 &  0.003 &  0.001 &  0.14 & $-$0.001 & $-$0.000 &  0.002 & $-$0.000 &  0.15 & $-$0.002 & $-$0.002 & $-$0.000 & $-$0.003 \\
 0.16 & $-$0.004 & $-$0.001 &  0.000 & $-$0.004 &  0.17 & $-$0.007 &  0.000 &  0.003 & $-$0.006 &  0.18 & $-$0.012 & $-$0.003 &  0.002 & $-$0.008 \\
 0.19 & $-$0.010 & $-$0.006 & $-$0.001 & $-$0.007 &  0.20 & $-$0.005 & $-$0.004 & $-$0.003 & $-$0.006 &  0.21 & $-$0.006 & $-$0.002 & $-$0.003 & $-$0.004 \\
 0.22 & $-$0.010 & $-$0.008 &  0.000 & $-$0.004 &  0.23 & $-$0.016 & $-$0.014 & $-$0.000 & $-$0.005 &  0.24 & $-$0.017 & $-$0.011 & $-$0.006 & $-$0.006 \\
 0.25 & $-$0.004 & $-$0.002 & $-$0.006 & $-$0.004 &  0.26 &  0.008 &  0.001 & $-$0.001 & $-$0.001 &  0.27 &  0.006 & $-$0.002 & $-$0.000 & $-$0.002 \\
 0.28 & $-$0.002 & $-$0.003 & $-$0.002 & $-$0.003 &  0.29 & $-$0.006 & $-$0.001 & $-$0.003 & $-$0.001 &  0.30 & $-$0.006 & $-$0.002 & $-$0.003 &  0.001 \\
 0.31 & $-$0.002 & $-$0.003 &  0.000 &  0.000 &  0.32 &  0.002 &  0.000 &  0.003 &  0.000 &  0.33 &  0.001 &  0.002 &  0.002 &  0.001 \\
 0.34 & $-$0.005 & $-$0.000 &  0.000 &  0.000 &  0.35 & $-$0.007 & $-$0.001 & $-$0.001 & $-$0.001 &  0.36 & $-$0.003 &  0.000 & $-$0.002 & $-$0.000 \\
 0.37 &  0.002 & $-$0.001 & $-$0.002 &  0.001 &  0.38 &  0.002 & $-$0.003 &  0.000 & $-$0.000 &  0.39 &  0.006 & $-$0.004 &  0.001 & $-$0.002 \\
 0.40 &  0.010 & $-$0.002 & $-$0.000 & $-$0.003 &  0.41 &  0.006 & $-$0.000 & $-$0.002 & $-$0.004 &  0.42 &  0.004 &  0.002 & $-$0.001 & $-$0.004 \\
 0.43 &  0.017 &  0.006 &  0.002 & $-$0.003 &  0.44 &  0.030 &  0.009 &  0.005 & $-$0.000 &  0.45 &  0.030 &  0.010 &  0.006 & $-$0.000 \\
 0.46 &  0.022 &  0.009 &  0.004 & $-$0.001 &  0.47 &  0.019 &  0.009 &  0.003 &  0.001 &  0.48 &  0.028 &  0.012 &  0.006 &  0.004 \\
 0.49 &  0.038 &  0.017 &  0.006 &  0.004 &  0.50 &  0.039 &  0.018 &  0.003 &  0.003 &  0.51 &  0.039 &  0.015 &  0.003 &  0.004 \\
 0.52 &  0.042 &  0.015 &  0.004 &  0.005 &  0.53 &  0.035 &  0.014 &  0.002 &  0.003 &  0.54 &  0.019 &  0.010 &  0.001 &  0.002 \\
 0.55 &  0.014 &  0.007 &  0.001 &  0.002 &  0.56 &  0.021 &  0.008 &  0.002 &  0.001 &  0.57 &  0.027 &  0.012 &  0.005 &  0.002 \\
 0.58 &  0.033 &  0.015 &  0.006 &  0.002 &  0.59 &  0.033 &  0.014 &  0.005 &  0.002 &  0.60 &  0.027 &  0.011 &  0.003 &  0.002 \\
 0.61 &  0.026 &  0.009 &  0.002 &  0.001 &  0.62 &  0.030 &  0.008 &  0.005 &  0.001 &  0.63 &  0.036 &  0.009 &  0.009 &  0.001 \\
 0.64 &  0.041 &  0.009 &  0.011 &  0.003 &  0.65 &  0.039 &  0.010 &  0.008 &  0.002 &  0.66 &  0.032 &  0.008 &  0.004 & $-$0.001 \\
 0.67 &  0.028 &  0.007 &  0.002 & $-$0.001 &  0.68 &  0.028 &  0.008 &  0.002 &  0.001 &  0.69 &  0.027 &  0.008 &  0.002 &  0.003 \\
 0.70 &  0.023 &  0.010 &  0.002 &  0.005 &  0.71 &  0.020 &  0.010 &  0.005 &  0.005 &  0.72 &  0.019 &  0.008 &  0.007 &  0.002 \\
 0.73 &  0.025 &  0.009 &  0.006 &  0.002 &  0.74 &  0.028 &  0.010 &  0.004 &  0.001 &  0.75 &  0.015 &  0.007 &  0.000 & $-$0.001 \\
 0.76 &  0.005 &  0.005 & $-$0.002 & $-$0.002 &  0.77 &  0.009 &  0.006 &  0.001 &  0.001 &  0.78 &  0.021 &  0.009 &  0.005 &  0.004 \\
 0.79 &  0.027 &  0.009 &  0.006 &  0.006 &  0.80 &  0.017 &  0.005 &  0.003 &  0.004 &  0.81 &  0.005 &  0.002 &  0.000 &  0.000 \\
 0.82 &  0.006 &  0.002 & $-$0.001 & $-$0.000 &  0.83 &  0.016 &  0.003 &  0.003 & $-$0.002 &  0.84 &  0.023 &  0.005 &  0.006 & $-$0.003 \\
 0.85 &  0.022 &  0.005 &  0.004 & $-$0.003 &  0.86 &  0.014 &  0.002 & $-$0.001 & $-$0.003 &  0.87 &  0.009 & $-$0.001 & $-$0.004 & $-$0.004 \\
 0.88 &  0.010 & $-$0.002 & $-$0.005 & $-$0.003 &  0.89 &  0.007 & $-$0.000 & $-$0.006 & $-$0.003 &  0.90 &  0.004 &  0.003 & $-$0.004 & $-$0.003 \\
 0.91 &  0.001 &  0.003 & $-$0.002 & $-$0.003 &  0.92 & $-$0.006 &  0.000 & $-$0.000 & $-$0.004 &  0.93 & $-$0.007 & $-$0.001 & $-$0.000 & $-$0.005 \\
 0.94 & $-$0.004 & $-$0.001 & $-$0.002 & $-$0.004 &  0.95 & $-$0.006 & $-$0.004 & $-$0.004 & $-$0.002 &  0.96 & $-$0.007 & $-$0.003 & $-$0.003 & $-$0.001 \\
 0.97 & $-$0.003 &  0.000 & $-$0.002 & $-$0.001 &  0.98 & $-$0.006 & $-$0.000 & $-$0.003 & $-$0.002 &  0.99 & $-$0.009 & $-$0.003 & $-$0.003 & $-$0.002 \\
 1.00 & $-$0.002 & $-$0.004 & $-$0.002 & $-$0.002 &  1.01 & $-$0.001 & $-$0.002 & $-$0.001 & $-$0.000 &  1.02 & $-$0.007 & $-$0.003 & $-$0.001 & $-$0.000 \\
 1.03 & $-$0.011 & $-$0.006 & $-$0.003 & $-$0.002 &  1.04 & $-$0.010 & $-$0.005 & $-$0.002 & $-$0.004 &  1.05 & $-$0.002 & $-$0.002 &  0.002 & $-$0.005 \\
 1.06 &  0.012 &  0.003 &  0.004 & $-$0.003 &  1.07 &  0.022 &  0.010 &  0.003 &  0.001 &  1.08 &  0.017 &  0.010 &  0.000 &  0.001 \\
 1.09 &  0.010 &  0.002 & $-$0.002 & $-$0.001 &  1.10 &  0.001 & $-$0.005 & $-$0.001 & $-$0.003 &  1.11 & $-$0.010 & $-$0.008 &  0.000 & $-$0.005 \\
 1.12 & $-$0.014 & $-$0.008 &  0.000 & $-$0.003 &  1.13 & $-$0.012 & $-$0.006 &  0.001 &  0.000 &  1.14 & $-$0.009 & $-$0.002 &  0.002 & $-$0.000 \\
 1.15 & $-$0.006 & $-$0.001 & $-$0.000 & $-$0.003 &  1.16 & $-$0.008 & $-$0.004 & $-$0.003 & $-$0.005 &  1.17 & $-$0.010 & $-$0.006 & $-$0.002 & $-$0.003 \\
 1.18 & $-$0.003 & $-$0.002 &  0.002 & $-$0.002 &  1.19 &  0.002 &  0.000 &  0.004 &  0.001 &  1.20 & $-$0.004 & $-$0.004 &  0.002 &  0.003 \\
 1.21 & $-$0.008 & $-$0.008 & $-$0.001 & $-$0.000 &  1.22 & $-$0.006 & $-$0.008 & $-$0.002 & $-$0.003 &  1.23 & $-$0.005 & $-$0.010 & $-$0.001 & $-$0.005 \\
\hline
\end{tabular}
\end{table*}


\begin{thebibliography}{}
\bibitem[Ahn et al (2012)]{Ahn etal12} {Ahn, C.~P., et al.} 2012, \textit{ApJS}, 203, 21
\bibitem[Berry et al.(2012)]{2012ApJ...757..166B} Berry, M., Ivezi{\'c}, {\v Z}., Sesar, B., et al.\ 2012, \apj, 757, 166
\bibitem[Blake et al.(2007)]{2007MNRAS.374.1527B} Blake, C., Collister, A., Bridle, S., \& Lahav, O.\ 2007, \mnras, 374, 1527
\bibitem[Chen et al.(2014)]{2014MNRAS.443.1192C} Chen, B.-Q., Liu, X.-W., Yuan, H.-B., et al.\ 2014, \mnras, 443, 1192
\bibitem[Covey et al.(2007)]{2007AJ....134.2398C} Covey, K.~R., Ivezi{\'c}, {\v Z}., Schlegel, D., et al.\ 2007, \aj, 134, 2398 
\bibitem[]{}Deng, L., Newberg, H., Liu, C., et al. 2012, RAA, 12, 735
\bibitem[Drew et al.(2005)]{2005MNRAS.362..753D} Drew, J.~E., Greimel, R., Irwin, M.~J., et al.\ 2005, \mnras, 362, 753
\bibitem[Eisenstein et al.(2001)]{2001AJ....122.2267E} Eisenstein, D.~J., Annis, J., Gunn, J.~E., et al.\ 2001, \aj, 122, 2267
\bibitem[High et al.(2009)]{2009AJ....138..110H} High, F.~W., Stubbs, C.~W., Rest, A., Stalder, B., \& Challis, P.\ 2009, \aj, 138, 110 
\bibitem[Ivezi{\'c} et al.(2004)]{2004AN....325..583I} Ivezi{\'c}, {\v Z}., Lupton, R.~H., Schlegel, D., et al.\ 2004, Astronomische Nachrichten, 325, 583 
\bibitem[Ivezi{\'c} et al.(2008)]{2008ApJ...684..287I} Ivezi{\'c}, {\v Z}., Sesar, B., Juri{\'c}, M., et al.\ 2008a, \apj, 684, 287
\bibitem[Ivezi{\'c} et al.(2007)]{2007AJ....134..973I} Ivezi{\'c}, {\v Z}., Smith, J.~A., Miknaitis, G., et al.\ 2007, \aj, 134, 973
\bibitem[Ivezic et al.(2008)]{2008arXiv0805.2366I} Ivezic, Z., Tyson, J.~A., Acosta, E., et al.\ 2008b, arXiv:0805.2366 
\bibitem[Juri{\'c} et al.(2008)]{2008ApJ...673..864J} Juri{\'c}, M., Ivezi{\'c}, {\v Z}., Brooks, A., et al.\ 2008, \apj, 673, 864 
\bibitem[Kaiser et al.(2002)]{2002SPIE.4836..154K} Kaiser, N., Aussel, H., Burke, B.~E., et al.\ 2002, \procspie, 4836, 154 
\bibitem[Landolt(1992)]{1992AJ....104..340L} Landolt, A.~U.\ 1992, \aj, 104, 340 
\bibitem[]{} Liu, X.-W., Yuan, H.-B, Huo, Z.-Y., et al., 2014, in Feltzing S., Zhao G., Walton N., Whitelock P., eds, Proc. IAU Symp. 298, Setting the scene for Gaia and LAMOST, Cambridge University Press, pp. 310-321, preprint (arXiv: 1306.5376)
\bibitem[Lee et al.(2008a)]{2008AJ....136.2022L} Lee, Y.~S., Beers, T.~C., Sivarani, T., et al.\ 2008a, \aj, 136, 2022
\bibitem[Lee et al.(2008b)]{2008AJ....136.2050L} Lee, Y.~S., Beers, T.~C., Sivarani, T., et al.\ 2008b, \aj, 136, 2050 
\bibitem[Padmanabhan et al.(2008)]{2008ApJ...674.1217P} Padmanabhan, N., Schlegel, D.~J., Finkbeiner, D.~P., et al.\ 2008, \apj, 674, 1217 
\bibitem[Padmanabhan et al.(2007)]{2007MNRAS.378..852P} Padmanabhan, N., Schlegel, D.~J., Seljak, U., et al.\ 2007, \mnras, 378, 852 
\bibitem[Planck Collaboration et al.(2013)]{2013arXiv1312.1300P} Planck Collaboration, Abergel, A., Ade, P.~A.~R., et al.\ 2013, arXiv:1312.1300 
\bibitem[Schlafly \& Finkbeiner(2011)]{2011ApJ...737..103S} Schlafly, E.~F., \& Finkbeiner, D.~P.\ 2011, \apj, 737, 103
\bibitem[Schlafly et al.(2012)]{2012ApJ...756..158S} Schlafly, E.~F., Finkbeiner, D.~P., Juri{\'c}, M., et al.\ 2012, \apj, 756, 158 
\bibitem[Schlafly et al.(2010)]{2010ApJ...725.1175S} Schlafly, E.~F., Finkbeiner, D.~P., Schlegel, D.~J., et al.\ 2010, \apj, 725, 1175 
\bibitem[Schlegel et al.(1998)]{1998ApJ...500..525S} Schlegel, D.~J., Finkbeiner, D.~P., \& Davis, M.\ 1998, \apj, 500, 525
\bibitem[Skrutskie et al.(2006)]{2006AJ....131.1163S} Skrutskie, M.~F., Cutri, R.~M., Stiening, R., et al.\ 2006, \aj, 131, 1163  
\bibitem[Stetson(2000)]{2000PASP..112..925S} Stetson, P.~B.\ 2000, \pasp, 112, 925 
\bibitem[Stetson(2005)]{2005PASP..117..563S} Stetson, P.~B.\ 2005, \pasp, 117, 563 
\bibitem[Stubbs \& Tonry(2006)]{2006ApJ...646.1436S} Stubbs, C.~W., \& Tonry, J.~L.\ 2006, \apj, 646, 1436 
\bibitem[Wright et al.(2010)]{2010AJ....140.1868W} Wright, E.~L., Eisenhardt, P.~R.~M., Mainzer, A.~K., et al.\ 2010, \aj, 140, 1868  
\bibitem[Wu et al.(2011)]{2011RAA....11..924W} Wu, Y., Luo, A.-L., Li, H.-N., et al.\ 2011, Research in Astronomy and Astrophysics, 11, 924
\bibitem[]{} Xiang, M.-S., Liu, X.-W., Yuan, H.-B., et al.,2014, \mnras, submitted
\bibitem[Yanny et al.(2009)]{2009AJ....137.4377Y} Yanny, B., Rockosi, C., Newberg, H.~J., et al.\ 2009, \aj, 137, 4377
\bibitem[York et al.(2000)]{2000AJ....120.1579Y} York, D.~G., Adelman, J., Anderson, J.~E., Jr., et al.\ 2000, \aj, 120, 1579
\bibitem[]{} Yuan, H.-B., et al., 2014a, \mnras, submitted
\bibitem[]{} Yuan, H.-B., et al., 2014b, ApJ, submitted
\bibitem[]{} Yuan, H.-B., et al., 2014c, ApJ, to be submitted 
\bibitem[]{} Yuan, H.-B., et al., 2014d, ApJ, to be submitted
\bibitem{}{} Yuan, H.~B. \& Liu, X.~W. 2012, \mnras, 425, 1763
\bibitem[Yuan et al.(2013)]{2013MNRAS.430.2188Y} Yuan, H.~B., Liu, X.~W., \& Xiang, M.~S.\ 2013, \mnras, 430, 2188
\end{thebibliography}
\end{document}